\documentclass[a4paper]{article}
\usepackage[english]{babel}
\usepackage[utf8x]{inputenc}
\usepackage[T1]{fontenc}
\usepackage{authblk}
\usepackage{multicol}
\usepackage{algorithm}
\usepackage{comment}
\usepackage{arydshln} 
\usepackage[noend]{algpseudocode}
\usepackage[colorinlistoftodos,backgroundcolor=lightgray]{todonotes}

\usepackage[a4paper,top=3cm,bottom=2cm,left=3cm,right=3cm,marginparwidth=1.75cm]{geometry}

\usepackage{amsmath}
\usepackage{graphicx}
\usepackage{soul}
\usepackage{setspace}
\title{Social Stability and Extended Social Balance  - \\ Quantifying the 
Role of Inactive Links in Social Networks }
\author[1,2]{Andres M.~Belaza}
\author[1]{Jan Ryckebusch}
\author[2,3,4]{Aaron Bramson}
\author[1]{Corneel Casert}
\author[2]{Kevin Hoefman}
\author[2]{Koen Schoors}
\author[1,2]{Milan van den Heuvel}
\author[1,2]{Benjamin Vandermarliere}
\affil[1]{Department of Physics and Astronomy, Ghent University}
\affil[2]{Department of General Economics, Ghent University}
\affil[3]{Laboratory for Symbolic Cognitive Development, RIKEN Brain Science Institute}
\affil[4]{Department of Software and Information Systems, University of North Carolina Charlotte}
\begin{document}
\maketitle
\begin{abstract}
Structural balance in social network theory starts from signed networks with active relationships (friendly or hostile) to establish a hierarchy between four different types of triadic relationships. The lack of an active link also provides information about the network. To exploit the information that remains uncovered by structural balance, we introduce the inactive   relationship that accounts for both neutral and nonexistent ties between two agents. This addition results in ten types of triads, with the advantage that the network analysis can be done with complete networks. To each type of triadic relationship, we assign an energy that is a measure for its average occupation probability.  Finite temperatures account for a persistent form of disorder in the formation of the triadic relationships. We propose a Hamiltonian with three interaction terms and a chemical potential (capturing the cost of edge activation) as an underlying model for the triadic energy levels. Our model is suitable for empirical analysis of political networks and allows to uncover generative mechanisms. It is tested on an extended data set for the standings between two classes of alliances in a massively multi-player on-line game (MMOG) and on real-world data for the relationships between countries during the Cold War era.  We find emergent properties in the triadic relationships between the nodes in a political network. For example, we observe a  persistent hierarchy between the ten triadic energy levels across time and networks. In addition, the analysis reveals consistency in the extracted model parameters and a universal data collapse of a derived combination of global properties of the networks.  We illustrate that the model has predictive power for the transition probabilities between the different triadic states.
\end{abstract}

\section{Introduction}

Balance theory \cite{heider1946,cartwright1956} captures a lot of the emergent phenomena in the relationships observed in political networks. Principles like ``a friend of a friend is also a friend''  and ``an enemy of my enemy is my friend'' are key driving forces in those networks.  The original formulation of balance theory by Heider~\cite{heider1946} and its extension to graphs by Cartwright~\cite{cartwright1956} is based on active relationships that can be friendly (``$+$'') or  unfriendly (``$-$''). The four types of emerging triadic relationships are categorized in two stable (=~balanced) and two unstable (=~unbalanced) ones, whereby one anticipates an overall tendency to create more balanced triads. Balanced triads $[+++]$ and $[+--]$ have an even number of ``$-$'' edges. Unbalanced $[++-]$ and $[---]$ triads, however, are a key ingredient in real-life political networks. Balance theory has found applications in many branches of sciences including psychology \cite{heider1946}, studies of international networks \cite{Hart1974, Doreian2015, Estrada2014, Galam1996, bramson2017measuring}, sociology \cite{Facchetti2011,Davis1967,Antal2005,Singh2016,SZELL2010} and ecology \cite{Saiz2017}. 

As balance theory introduces correlations between the edge attributes in triads, in a physics framework~\cite{Marvel2009, Belaza2017, Lee2017, DIEP2017} it maps onto a system with predominant three-edge  interactions. Associating the existence of unbalanced triads to the occurrence of a non-vanishing excitation energy, the principles of social balance can be mapped onto a model with variations in an energy landscape. Marvel et al.~\cite{Marvel2009} investigated the energy landscape of a social-balance inspired system and stressed the occurrence of so-called jammed states: network configurations with a finite number of unbalanced triads that cannot evolve to a fully balanced situation. A thermodynamical formulation of balance theory inspired by the Ising model is introduced in~\cite{Lee2017} by Lee and collaborators. With the aid of Markov chain Monte Carlo sampling, one determines probabilities to find a certain number of unbalanced triads. Connections between the simulations and real-world networks are established through the introduction of a social temperature that controls the fraction of unbalanced triads. 

Conflict and the role of coalitions in networks have been addressed from the physics perspective. Galam~S. in~\cite{Galam1996} proposed a model for international relationships inspired by the Ising model. The countries can belong to two coalitions and this information is encoded in their spin value $s_{i} \in \left\{ -1, +1 \right\}$. The total energy to minimize is $ -\sum _{i < j} J_{ij} s_{i} s_{j}$ whereby the signs of the $J_{ij}$ are inferred from the history of the inter-country relationships. 
The system works as a fully connected, weighted and signed network, as the $J_{ij}$ can adopt a real value. The system reaches the ground state if friendly countries $(J_{ij} > 0)$ are both members of  the same alliance $(s_i = s_j)$ and hostile countries $(J_{ij}< 0)$ are in different coalitions. This system displays a certain degree of frustration for triads with hostile ties between the nodes. {Vinogradova and Galam in~\cite{Vinogradova2014} extended this model with multiple layers and used it to study the internal dynamics of the European Union. Diep, et al.~in~\cite{DIEP2017,Kaufman2018} propose a model that describes the interaction between groups. In this model, the preference or attitude is encoded as a continuous spin, and the energy is defined as the interaction between these attitudes inside and between groups.}

Next to a statistical physics framework, other approaches have been proposed to account for the occurrence of  unbalanced triads. For example, Deng~et al.~\cite{DENG2016} and Du et al.~\cite{DU2018} suggested that  the counteracting forces of homophily and heterophobia inhibit the network to reach a fully balanced state. {Various alternative measures of balance have been proposed (for example, see~\cite{Estrada2014}) and a comparitive study for different networks can be found in Kirkley et al.~\cite{Kirkley2018}.}

Obviously not all relationships can be categorized as positive or negative. In international relationship studies, for example, the Swiss Confederation is an example of an agent that maintains neutrality.
An extension of balance theory to include neutrals is described in ~\cite{Gawroski2005}. This work introduces continuous edge values in the range $[-1, + 1]$  and reformulates balance theory as a set of dynamical equations. Thereby, $0$-valued links can be interpreted as the representation of a neutral inter-nodal relation. Under very specific circumstances, the fully connected network is found to evolve to Heider's balanced situation (or ``utopia''). 

Lerner~\cite{Lerner2016} has investigated the correlation between the geographical parameters (attributes of the countries) and the status of the inter-country relationships between $1885$ to $2001$.  A conclusion was that social balance does not predict well the value of a possible link, but is only able to predict the value of a link conditional on its activation. This means that given that there is an active (``$+$'' or ``$-$'') relationship, balance theory can predict its sign.

In a previous publication~\cite{Belaza2017} we have shown that structural balance can be mapped onto a Boltzmann-Gibbs type of Hamiltonian model. The model assigns a specific energy to the  four types of triads from structural balance and allows one to quantify their relative occupation probabilities. The proposed Hamiltonian has three-edge, two-edge and one-edge interactions that induce correlations between the edge values. 
In line with social balance, the three-edge term has the greatest coupling strength. Here we build on the work of ~\cite{Belaza2017} and take also into account the effect of inactive (neutral or nonexistent) edges in political networks. In this way, we add a dynamical aspect to the model and account for the temporal evolution of the network's topology by means of the activation or deactivation of the links.
 This implies that we go beyond the restrictions of signed networks and introduce inactive (``0'') as a possible status of relationship between two agents in the system. This extension makes the network complete which facilitates analytical derivations in the mean-field approximation for example. In addition, triads with only active edges usually represent but a few percent of all possible triads in the political network. We anticipate that the lack of an active link is a source of information about the structure of the network and of the political relationships. Our proposed methodology can unravel that information. We test the proposed extension of balance theory against two different datasets. The first dataset is the international relationship network for the Cold War era (1949-1993)~\cite{COW}. As outlined in \cite{Belaza2017}, we constructed that network and its time evolution by combining data of military alliances and data of inter-state disputes \cite{AlliancesDatabase,MIDs}. The second dataset stems from a virtual world called EVE Online~\cite{EveOnline}. These detailed data provide the daily evolution of the standings between the alliances of players in EVE and cover a period of over a year.

In the forthcoming section~\ref{sec:data} we provide a description of the datasets used in this work. We then proceed (section~\ref{sec:hamiltoninan}) with introducing the Hamiltonian that corresponds with the proposed extended version of balance theory. We introduce some global properties that can be computed from the model (section~\ref{subsec:hamilgeneral}). We also provide a mean-field approximation for the proposed Hamiltonian (section~\ref{subsec:MeanField}) that can be used to compute those global properties. We continue with providing detailed theory-data comparisons for both static (section~\ref{subsec:staticprobabil}) and dynamic (section~\ref{subsec:dynamicproba}) properties of the studied political networks. 
Section~\ref{subsec:ResultsMeanField} is devoted to a comparison of the recorded network properties and the predictions from the mean-field approximation to the proposed Hamiltonian for the triadic states.

\section{Datasets for relationships in political networks}
\label{sec:data}

\begin{figure}[htb]
\begin{flushleft}
\centering
\includegraphics[width=0.49\textwidth]{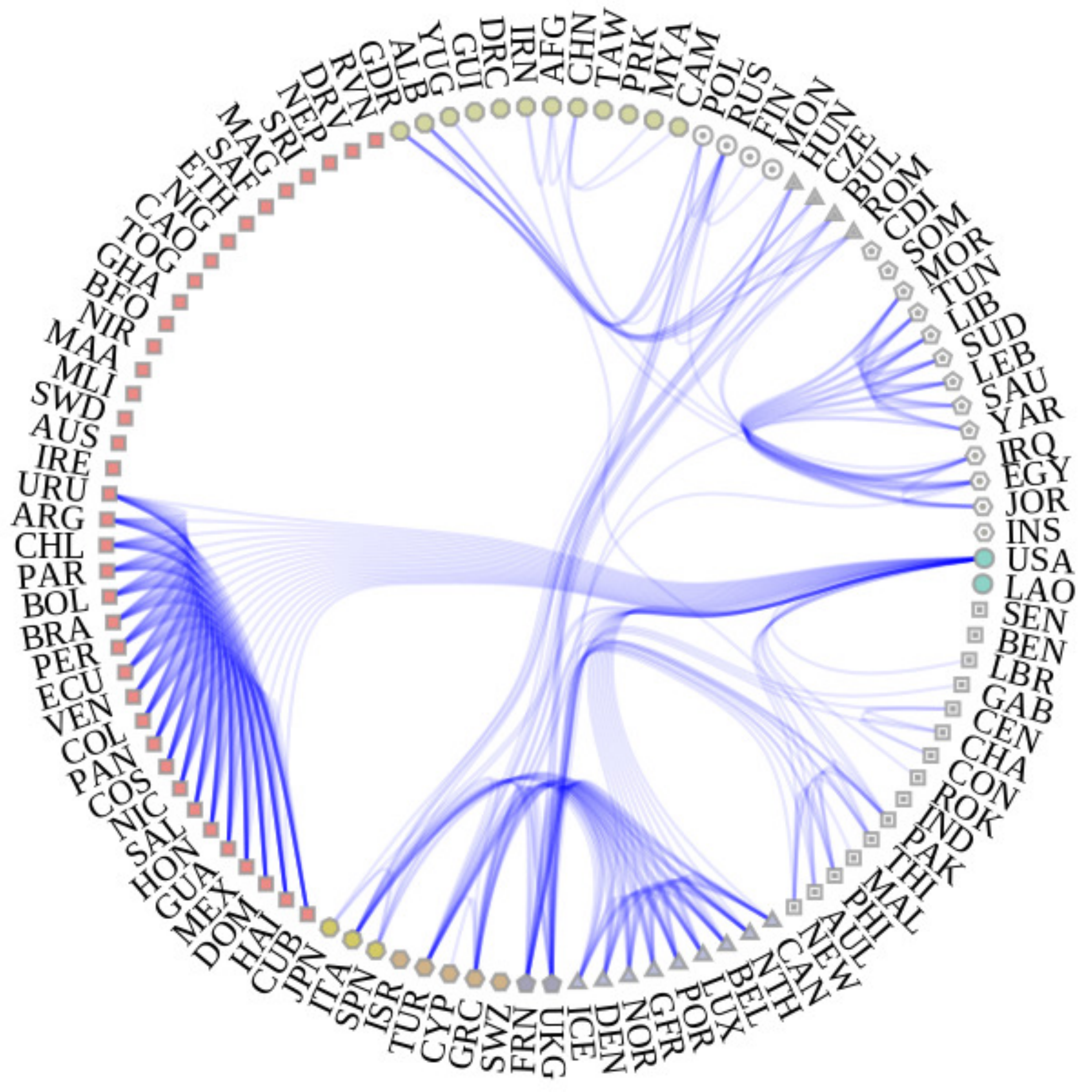}
\includegraphics[width=0.49\textwidth]{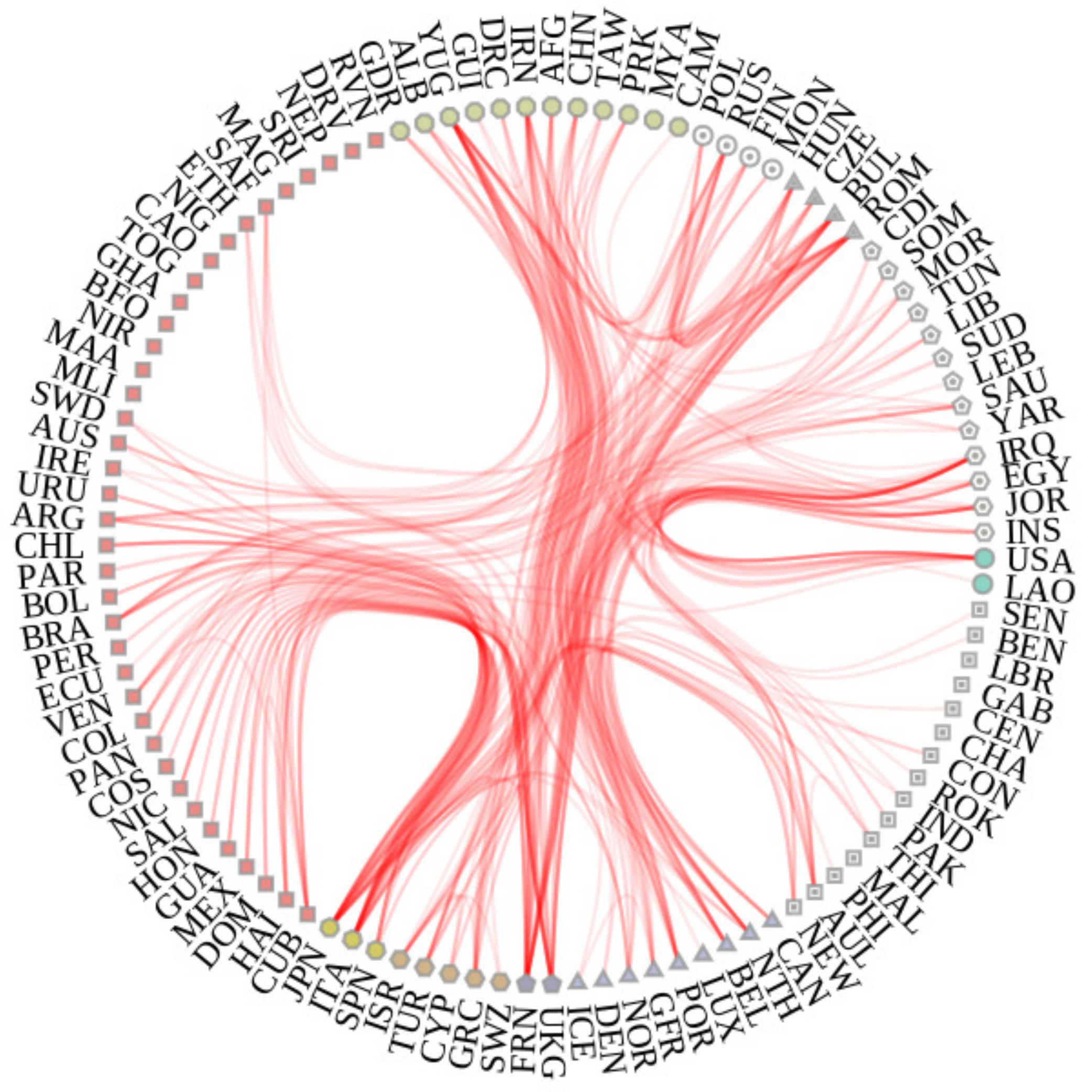}
\end{flushleft} 
\caption{ 
Prototypical example (1960) of the yearly Cold War alliance network. Each block (color-coded) is a group of alliances that have a similar position/play a similar role in the alliance network. For clarity, we disentangled the  positive (left panel) and negative links (right panel). The inactive links are not drawn and some newly established countries (mostly former colonies gaining independence) do not have any signed edges yet. The displayed hierarchical block structure was detected with the methodology described in ~\cite{Peixoto2015}. } 
\label{fig:NetExample}
\end{figure}

Throughout the paper we use two different datasets on political relationships to challenge our proposed extension of balance theory.

The first dataset reports on international relationships during the Cold War era (1949-1993) and was collected by the Correlates of War initiative~\cite{COW}. This dataset has records on international alliances \cite{AlliancesDatabase} as well as interstate disputes \cite{MIDs} which can range from commercial or military blocks, to threats and open war. For every year we construct a signed network from the bilateral records. Thereby, two allies share a positive link whereas two countries in a dispute share a negative link. In those situations where both are present, we consider the most recent one. A pair of countries is considered to have a neutral link when no alliances nor disputes between them are registered.
As an illustration, the international alliance network and its emerging structure can be seen in figure~\ref{fig:NetExample} for the year $1960$. 
One can e.g. distinguish a clear group of positively linked Latin-American countries with also a positive link to the USA.
For reasons of clarity, neutral links were not drawn. 
Further details on the dataset and the construction of the network are outlined in ~\cite{Belaza2017} together with an analysis of the resulting yearly networks. 

The second dataset was extracted from a virtual world called EVE Online~\cite{EveOnline}. EVE Online is a sandbox massive multi-player on-line game (MMOG) developed by CCP games. In this virtual world, more than 500,000 players trade, collaborate and fight in a futuristic galaxy. The players organize themselves in social structures called alliances with sizes between one and about 25,000 players. The alliances can conquer territory, where they can impose their own taxes, exploit mineral resources, and so on. Because the data comes from a virtual world, complete and accurate records of all alliance standings across time are available and advanced statistical analyses become feasible.
The dataset tracks the daily evolution of the standings between the alliances of players in the MMOG and covers a period of over a year.
The relations between the alliances represent an important aspect of the game as they impact a wide range of game-play experiences. 
The leadership of an alliance can explicitly and publicly set these relationships to friendly, hostile, neutral or undetermined. 
This is important because these standings affect how players from one alliance react to players from another alliance by facilitating the process of discriminating between friends, enemies, and others. Most alliances in EVE follow a ``Not Blue, Shoot It'' policy: friends are to be left in peace, while any other relationship means to shoot on sight. However, military, planed and coordinate actions between players have mainly enemy alliances as objectives.

For the entire data period ranging from March 2015 to April 2016, we use the standings data to construct two versions of the daily signed alliance relationship network.
One version consists of the alliances that have more than 200 members (so called “+200” alliances), the other of the alliances that hold sovereignty over at least one solar system (“SOV” alliances for short). 
These two classes of alliances are key to the political dynamics of the game.

We note that before one can construct the undirected signed network which is required by our framework, it is necessary to make the  sporadically asymmetrical standings symmetrical. The conversion rules for this symmetrization are the following: if at least one of the alliances considers the other an enemy, the link is negative. If this is not the case, and at least one of them thinks the relation is amicable, the link is positive.

The remaining combinations give rise to inactive links: neutral/neutral and neutral/not set. This symmetrization process impacts the degeneracy, i.e.~the total number of possible ways to create a specific type of triad. The degeneracies are listed in table~\ref{table:Levels}. The conversion rules are inspired by the rules of the game so that the status of the relationship between the alliances reflects the dynamics of EVE Online. 
For more details on the data and network construction, we refer the reader to the discussion in~\cite{Belaza2017}.

\section{A Hamiltonian approach to extended social balance}
\label{sec:hamiltoninan}
\subsection{Formalism}
\label{subsec:hamilgeneral}

In a previous work~\cite{Belaza2017}, we proposed a  Hamiltonian model with four emerging types of triads that can capture the general features of social balance, including the preference for balanced states and the importance of the three-edge forces in political graphs.  The simplest version of social balance only discriminates between balanced and unbalanced triads in the graph. 
In our approach, social balance is the result of ordering and disordering mechanisms. The ordering is induced by the fact that the balanced triads possess a lower energy than the unbalanced ones. Exogenous effects, like resource interests, geographical and socio-economic parameters, have a disordering impact and introduce unbalanced triads. We introduce ``social temperature'' as a global measure for the disordering effect.  

We suggest that the parameter $s_{ij}$ that quantifies the relation  between agents $i$ and $j$ can adopt three discrete values:  $s_{ij}=+1$ for friendship, $s_{ij}=-1$ for enmity, and $s_{ij}=0$ for a non-existent or a neutral relation. We propose to extend the Hamiltonian introduced in \cite{Belaza2017} with the possibility to activate and deactivate ties in the network. This can be achieved through the introduction of a chemical potential $\mu$, which is the energy cost of activating a link ($s_{ij}=0 \rightarrow s_{ij}=\pm 1$). $\mu$ is also the energy gain upon deactivating ($s_{ij}= \pm 1 \rightarrow s_{ij}=0$)  a link.  We refer systematically to $s_{ij}=0$ links as ``inactive'' ones.

The state of the political network consisting of $N$ agents is uniquely defined by the numbers 
\begin{equation}
\left\{ s_{ij} \right\} \equiv s_{12},s_{13},\ldots,s_{1N},s_{21}, s_{23}, \ldots s_{2N}, \ldots, s_{N1}, s_{N2}, \ldots, s_{NN-1} \; 
\label{eq:statesjan}.
\end{equation}
We consider undirected networks implying that $s_{ij} = s_{ji}$. The proposed extended Hamiltonian in the space of edge values reads
\begin{eqnarray}
\mathcal{H} \left(\left\{ s_{ij} \right\} \right) & = & \frac{1}{6} \sum_{i \ne j \ne k=1}^{N}  \biggl[  
\underbrace{- \; \alpha \; s_{ij} s_{ik} s_{jk}}_{\text{three-edge interaction}} 
\underbrace{- \; \gamma  \; \left( s_{ij}s_{ik}+s_{ij}s_{jk}+s_{jk}s_{ik} \right) }_\text{two-edge interaction} \biggr] 
\nonumber \\
& &  +  \frac{1}{2} \sum_{i \ne j=1}^{N}  \biggl[ \underbrace{\; \;    + \omega \; s_{ij}}_{\text{one-edge interaction  }} \;
\underbrace{ \; + \; \mu \;  s_{ij}^2 }_{\text{chemical potential  }} \biggr] \; ,
\label{eq:Hamiltonian}
\end{eqnarray}
with $s_{ij} \in \left\{ -1,0,+1 \right\}$ and $N$ the total number of agents in the network. In practice, the sum $\sum_{i \ne j \ne k=1}^{N}$ runs over all possible triads in the network. The factor $\frac{1}{6}$ accounts for the fact that the triad $T_{ijk}$ defined by the agents $\{i,j,k\}$ appears six times in the summation. Similarly, the factor  $\frac{1}{2}$ accounts for the fact that each edge occurs twice in the summation $ \sum_{i \ne j}^{N}$.  

\begin{figure}[htb]
\begin{flushleft}
\includegraphics[width=0.85\textwidth]{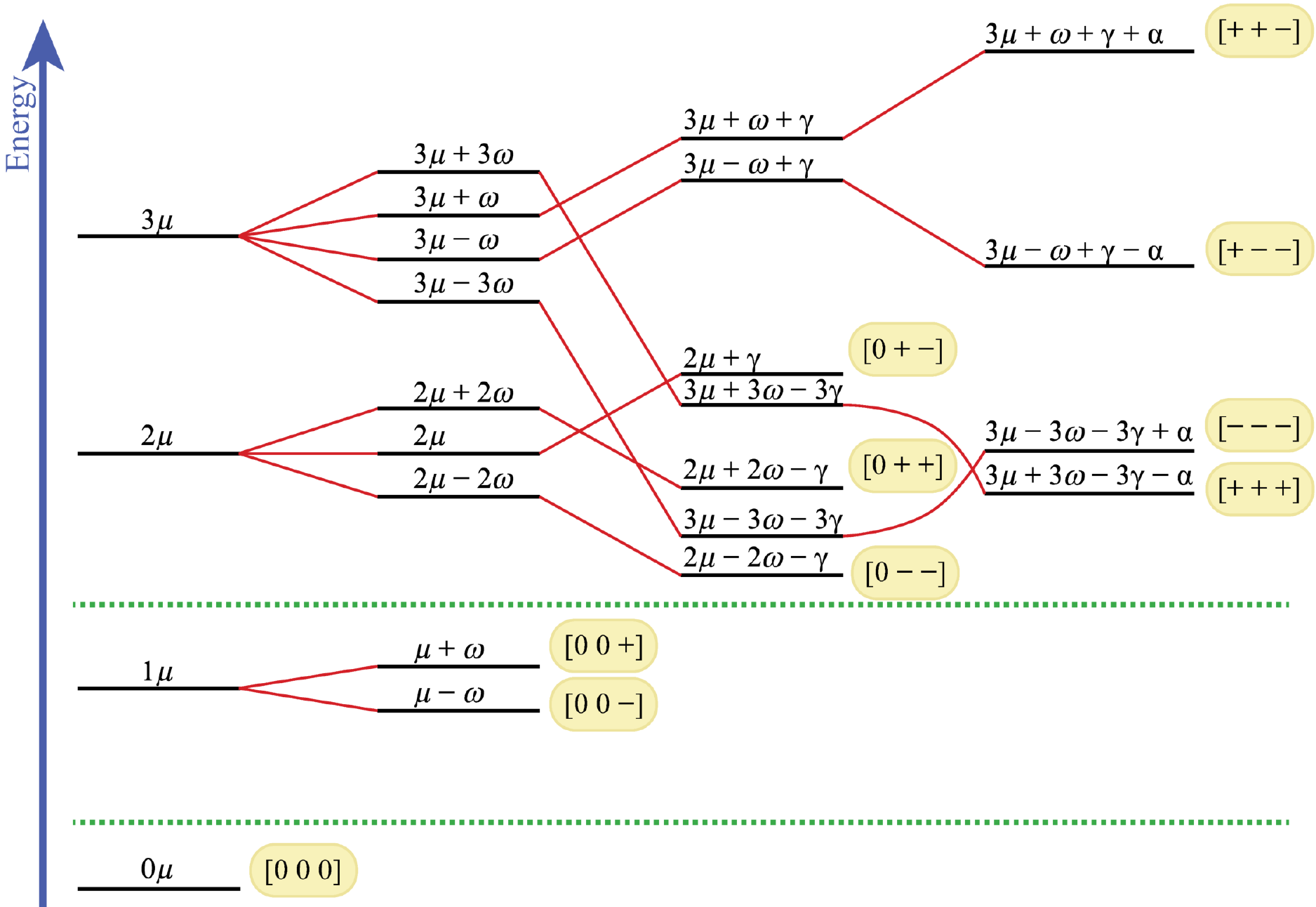}
\end{flushleft} 
\caption{The different triadic energy levels that correspond with the Hamiltonian of equation~(\ref{eq:Hamiltonian}). The order of occurrence of the different energy levels depends on the parameters. 
The displayed triadic energies are for $\{ \alpha = 0.40, \gamma= 0.36, \omega = 0.10, \mu = 1.0\}$, numbers that are in the ballpark of what we extract from data (see table ~\ref{table:Adjusted Parameters1}).}  
\label{fig:Diagram}
\end{figure}

\begin{table}[!htb]
	\begin{center} \renewcommand{\arraystretch}{1.2}
\resizebox{\textwidth}{!}{%
    \begin{tabular}{ l l l l l } 	
       \hline \hline
            &
            & \multicolumn{1}{c}{Associated energy}
            & \multicolumn{1}{c}{Degeneracy}
            & \multicolumn{1}{c}{Degeneracy} \\
            \multicolumn{1}{c}{Type of triad}
            & \multicolumn{1}{c}{Symbol $\sigma$}   
            & \multicolumn{1}{c}{from Hamiltonian (\ref{eq:Hamiltonian})}
            & \multicolumn{1}{c}{(Cold War)} 
            & \multicolumn{1}{c}{(EVE)} \\ \hline 
Highly frustrated 	& A: $[++-]$ 	& $E_A = +\alpha+\gamma+\omega+3\mu$ 	& $g_G (E_A)=24$ & $g_T (E_A)=525$\\ 
Lowly frustrated 	& B: $[---]$ 	& $E_B = +\alpha-3\gamma-3\omega+3\mu$ 	& $g_G (E_B)=8$ & $g_T (E_B)=343$\\ 
Lowly balanced 		& C: $[+--]$ 	& $E_C = -\alpha+\gamma-\omega+3\mu$ 	& $g_G (E_C)=24$ & $g_T (E_C)=735$\\ 
Highly balanced 	& D: $[+++]$ 	& $E_D = -\alpha-3\gamma+3\omega+3\mu$ 	& $g_G (E_D)=8$ & $g_T (E_D)=125$\\ 
\hline
Singly inactive v1	& E: $[0\;+-]$ 	& $E_E = +\gamma+2\mu$ 	& $g_G (E_E)=24$ & $g_T (E_E)=840$\\
Singly inactive  v2	& F: $[0\;++]$ 	& $E_F = -\gamma+2\omega+2\mu$ 	& $g_G (E_F)=12$ & $g_T (E_F)=300$\\
Singly inactive v3	& G: $[0\;--]$ 	& $E_G = -\gamma-2\omega+2\mu$ 	& $g_G (E_G)=12$ & $g_T (E_G)=588$\\
\hdashline
Doubly inactive v1	& H: $[0\;\;0\;+]$ 	& $E_H = +\omega +\mu $ 	& $g_G (E_H)=6$ & $g_T (E_H)=240$ \\                    
Doubly inactive v2 	& I: $[0\;\;0\;-]$ 	& $E_I = -\omega +\mu$ 	& $g_G (E_I)=6$ & $g_T (E_I)=336$ \\
\hdashline
Fully inactive		& J: $[0\;\;0\;\;0]$ 	& $E_J = 0$ 	& $g_G (E_J)=1$ & $g_T (E_J)=64$\\
            \hline \hline
			\end{tabular}}
		\caption{The energies and degeneracies corresponding with the ten types of triadic relationships in a complete political network with edges that can adopt the values ``$+$'', ``$-$'' and ``$0$''. We refer to ``$+$'' and ``$-$'' edges as active ones, and to ``$0$'' edges as inactive ones. The triads can be separated in groups with 0, 1, 2 and 3 active edges. } 
		\label{table:Levels}
	\end{center}
\end{table}

The triads can be in 10 different states $\sigma \in \left\{ A,B,C,D,E,F,G,H,I,J \right\} $ and the Hamiltonian~(\ref{eq:Hamiltonian}) determines the corresponding energies $E_\sigma$ (see table ~\ref{table:Levels} and figure ~\ref{fig:Diagram}). The Hamiltonian~(\ref{eq:Hamiltonian}) bears resemblance with the one from the Blume-Capel model \cite{Blume1966,CAPEL1966, Blume1971, SANTOS2018}, with the addition of a three-edge interaction related to standard social balance. The corresponding strength parameter $\alpha$ is anticipated to be positive.  The two-edge interaction term  reflects the tendency to ``homogenize'' the active relations in the triad. It introduces a fine-splitting of magnitude $4 \gamma$ in the energy spectrum of the triads with three active edges depending on whether the balanced or unbalanced triad is symmetric or not (see table ~\ref{table:Levels} and figure ~\ref{fig:Diagram}). The creation of a ``$+$'' or a ``$-$'' link between two agents in a political network comes with a ``cost'', and that is why we anticipate positive values of the chemical potential $\mu$. This is represented by the last term in the Hamiltonian of equation~(\ref{eq:Hamiltonian}), where  $s_{ij}^2$ has a value of 1 if a link is active, and 0 otherwise. The proposed Hamiltonian model allows one to describe a political system as a fully connected network, where the links must be one of positive, negative or neutral/nonexistent. An alternative and more dynamic picture is that of a political system as an  incomplete and irregular network with a varying number of positive or negative edges that are deactivated to and activated from neutral/nonexistent. 

  With the aid of the Hamiltonian of equation~(\ref{eq:Hamiltonian}), the probability of finding the political network in a state $\left\{ s_{ij} \right\}$ is determined by an expression of the type
\begin{equation}
p \left( \left\{ s_{ij} \right\} \right)  = \frac 
{e^{-\beta \mathcal{H} \left(\left\{ s_{ij} \right\} \right)}} 
{\sum_{s_{ij}=\pm 1,0} e^{-\beta \mathcal{H} \left(\left\{ s_{ij} \right\} \right) }}.
\label{eq:ProbabilityOfState}
\end{equation}
Hereby, we have introduced the positive-valued inverse temperature $\beta$ in the standard definition used in physics. Accordingly, the value of the temperature is a measure for the coupling of the political network with its environment. At vanishing temperature, all $T_{ijk}$ reside in the $[0\; 0\; 0]$ state. 

\begin{figure}[htb]
\centering
\includegraphics[width=1.0\textwidth]{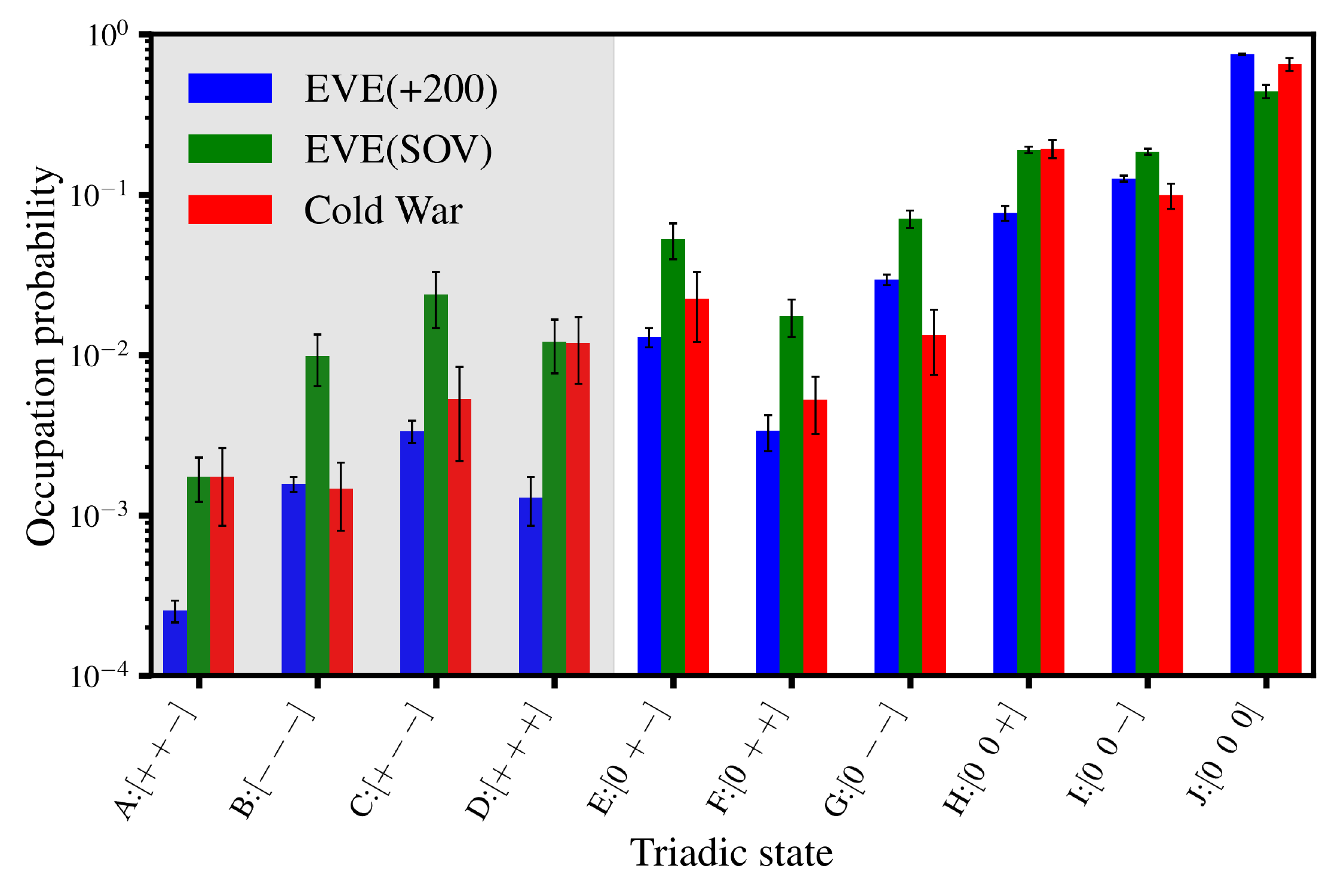}
\caption{Time-averaged occupation probabilities for each of the 10 types of triads and for each of the three datasets. The shaded area corresponds with the four triads of standard structural balance theory.}
\label{fig:Histogram_ocupation_probabilities}
\end{figure}

In figure ~\ref{fig:Histogram_ocupation_probabilities} we display the time-averaged occupation probabilities for all 10 types of triads for the three  political networks introduced in section~\ref{sec:data}.  A similar type of hierarchy for the occupation probabilities is found for the EVE and Cold War data. Roughly speaking, the larger the amount of active edges in a triad, the less probable its occurrence. There is a clear difference in the occupation probabilities for the triads with no activated edge and one activated edge. More than 40\%  of all triads are in a $[0\; 0\; 0]$ state.  The two types of triads with one activated edge $[0\; 0 \; \pm]$ represent each about 20\% of the total number of triads.  The seven types of triads with three activated edges (structural balance) and two activated edges are characterized by occupation probabilities at the single-digit percent level. This implies that the four types of triads with three activated edges that define standard structural balance, represent at most a couple of percent in the total population of triads. In section~\ref{subsec:staticprobabil}, the full time series of the empirical occupation probabilities of figure ~\ref{fig:Histogram_ocupation_probabilities}  will be used to learn something about the generative mechanisms in the political network. This will be done by optimizing the values of the strength parameters in the Hamiltonian of equation~(\ref{eq:Hamiltonian}) so that they can capture the observations.

Given a measure $M$ of the political network, its expectation value is given by
\begin{equation}
\left<M \right> =  \sum_{s_{ij}=\pm 1,0} 
M \left( \left\{ s_{ij} \right\} \right) 
p \left( \left\{ s_{ij} \right\} \right) .
\label{eq:MeanValueM}
\end{equation}

For a given combination of strength parameters $\{ \alpha,\gamma,\omega \}$ and an inverse temperature $\beta$, the equation~(\ref{eq:statesjan}) determines the probabilities $p \left( \left\{ s_{ij} \right\} \right)$ for all possible network structures and the expectation values $ \left< M \right>$. 

We propose two global measures that are characteristic for the political network. The first measure is related to the expectation value of the edges in the network   

\begin{equation}
L \equiv \left<s_{ij}\right> \hspace{0.05\textwidth} \left( -1 \le L \le +1 \right) \; ,
\label{eq:average_rela}
\end{equation}
and can be interpreted as the ``average magnetization'' in the system.  The quantity $L$ determines the average relationship between the agents in the network. The second quantity $A$ is indicative for the fraction of the links that are part of an active ($s_{ij}=\pm 1$) relationship 

\begin{equation}
A \equiv \left<s_{ij}^2\right> \hspace{0.05\textwidth} \left( 0 \le A \le +1 \right)  \; .
\label{eq:activation}
\end{equation}

We will refer to the $A$ as the ``average activation'' in the network. In our case the edge values are one of [-1,0,1] so the use of $s_{ij}^2$ acts like the absolute value, but we follow the BC model \cite{Blume1966,CAPEL1966} here because the squared value has methodological advantages in the derivations below. The value of $A$ is a measure of the tendency to create active (``$+$'' or ``$-$'') edges in the network.  Obviously, there is a strong correlation between the value of $A$, the value of the chemical potential $\mu$ and the temperature. Without loss of generality, in what follows we set the chemical potential as the unit of energy ($\mu \equiv 1$) and the zero of energy at the energy of the fully inactive triad $[0\;0\;0]$ ($E_{J}=0)$.

\subsection{Mean-field approximation}
\label{subsec:MeanField}

We now apply a mean-field approximation \cite{Kadanoff2009,Costabile2012} to the Hamiltonian of equation~(\ref{eq:Hamiltonian}). This is done in the standard way, by isolating the fundamental degrees of freedom (for our purposes, the edges $s_{ij}$) and setting them in their environment that is due to their  couplings with the other edges in the political network. To this end, we rewrite the Hamiltonian of equation~(\ref{eq:Hamiltonian}) as 

\begin{equation}
6 \mathcal{H}  =  \sum_{i \ne j} \biggl[ \sum_{k \ne i, k \ne j} \biggl[-\alpha s_{ij}   s_{ik}s_{jk} 
-\gamma s_{ij}   \left( s_{ik}+s_{jk} \right) 
-\gamma  s_{ik} s_{jk} \biggr]
+ 3\omega s_{ij}   + 3\mu s_{ij}^2  \biggr] \; .
\end{equation}
Proceeding in the standard way by writing the edges $s_{ik}$ and $s_{jk}$ as the sum over their expectation value $L = \left< s_{ij} \right> = \left< s_{jk} \right>$ and the deviation $\Delta_{ij}$ through $s_{ij} = L + (s_{ij}-L) = L + \Delta_{ij}$, one obtains

\begin{eqnarray}
6 \mathcal{H} & = &  \sum_{i \ne j} \biggl[ \sum_{k \ne i, k \ne j} \biggl[  
-\alpha s_{ij} \left(  L^2 + L(\Delta_{ik}+\Delta_{jk})+\Delta_{ik}\Delta_{jk} \right) 
-\gamma s_{ij} \left(  2L + \Delta_{ik}+\Delta_{jk} \right) 
\nonumber \\ 
& - & \gamma \left(  L^2 + L(\Delta_{ik}+\Delta_{jk})+\Delta_{ik}\Delta_{jk} \right) \biggr]
 + 3\omega s_{ij}  + 3\mu s_{ij}^2   \biggr] \; .
\end{eqnarray}
Expanding up to zeroth order in the deviations $\Delta_{ab}$ from the mean $L$, we have that 

\begin{eqnarray}
\mathcal{H} & = & \frac {1} {2} \sum_{i \ne j} \mathcal{H}^{ij}_{MF} + \theta( \Delta )
=  \frac {1} {2} \biggl[ \sum_{i \ne j} - \frac {1} {3} \alpha s_{ij} L^2 (N-2) - \frac {1} {3} \gamma s_{ij} 2L  (N-2) - (N-2) \gamma L^2   
\nonumber \\ 
& + & \omega s_{ij}  +\mu s_{ij}^2  \biggr] + \theta( \Delta ) \; .
\label{eq:MeanField0}
\end{eqnarray}
This means that in the mean-field approximation the Hamiltonian corresponding with an edge $s_{ij}$ adopts the form

\begin{eqnarray}
\mathcal{H}_{MF}^{ij} & = & \mu s_{ij}^2 +  \biggl(\omega -  \frac {2 \gamma} {3} (N-2) L -\frac {\alpha} {3} (N-2) L^2 \biggr) s_{ij} +  \frac {\gamma} {3} (N-2) L^2   
\nonumber \\
& = &  c_2 (\mu) s_{ij}^2 + c_1 (N, L, \alpha, \gamma, \omega) s_{ij}  
+ c_0 (N, L, \gamma )\; .
\label{eq:MeanField}
\end{eqnarray}
Remark that $c_2$ is uniquely defined by the chemical potential, whereas $c_1$ depends on the strength parameters of the three interaction types in the Hamiltonian of equation~(\ref{eq:Hamiltonian}).

The partition function corresponding with one edge can be readily obtained in the mean-field approximation by performing the summation over all possible $s_{ij}$ in 
$\mathcal{H}_{MF}^{ij}$. One obtains  
\begin{equation}
Z^{ij}_{MF} =  (e^{\beta c_2} + 2\cosh \beta c_1) e^{-\beta (c_2+c_0)} \; .
\end{equation}

The values of the measures 
$L$ and $A$ of Eqs.~(\ref{eq:average_rela}) and (\ref{eq:activation}) can be directly computed from the partition function 
\begin{equation}
L= 
\left<s_{ij}\right>  =  
\frac
{ - \partial \ln Z^{ij}_{MF}}
{ \partial (\beta c_1)} = 
\frac{-2\sinh \beta c_1 (N, L, \alpha, \gamma, \omega)}{(2\cosh \beta c_1 (N, L, \alpha, \gamma, \omega) +e^{\beta c_2 (\mu)})} \; ,
\label{eq:MFMagnetitation}
\end{equation}
\begin{equation}
A= 
\left<s_{ij}^2\right> =   \frac 
{ - \partial \ln Z^{ij}_{MF}}  { \partial \left( \beta c_2 \right) } = 
\frac{2\cosh \beta c_1 (N, L, \alpha, \gamma, \omega) }{(2\cosh \beta c_1 (N, L, \alpha, \gamma, \omega) +e^{\beta c_2 (\mu) })} \; .
\label{eq:MFExistence}
\end{equation}

The equation for the $L$ is a self-consistent equation.  The self-consistent equation for the magnetization of the infinite-range Ising Hamiltonian can be retrieved from equation~(\ref{eq:MFMagnetitation}) in the limit of $\mu \rightarrow -\infty$ and vanishing three-edge interactions.  

Whereas the average activation $A$ is rather insensitive to changes in the strength parameters, distinct regimes for the average magnetization $L$ can be identified.
At low temperatures ($ \beta ^{-1} \ll \mu$) there is not enough energy to activate the links and one has $L \rightarrow 0$ and $ A \rightarrow 0$. 
For an intermediate range of temperatures ($\beta ^{-1} \approx \mu$), links start to get activated and a non-zero magnetization dependent on the parameters $\alpha$, $\gamma$ and $\omega$ emerges.  For combinations of parameter values that result in a positive (negative) $c_1$, the average magnetization is negative (positive).  At high temperatures ($ \beta ^{-1} \gg \mu$)  all triadic states are equiprobable. Accordingly, one has $L \rightarrow 0$ and $ A \rightarrow \frac{2}{3}$, as there are two active (``$+$'' and ``$-$'') and one inactive (``$0$'') possibilities for the edge attributes.

As the system measures $L$ and $A$ are rather simple and common measures, the mean-field results restate some well-known features of the dynamics of political networks.
Indeed, in the absence of incentives to create active ties, the network will be fully inactive.  As soon as some incentives appear, the relations start to form and the links get activated.  When these incentives are abundant and omnipresent, the network behaves like a random network. Many systems are expected to fall in the intermediate range where the incentives are present and create some form of order. The derived mean-field expressions  (\ref{eq:MFMagnetitation}) and (\ref{eq:MFExistence}) give rise to some non-trivial and intriguing relation between the system measures $L$ and $A$ and other system properties. This prediction will be the topic of investigation in section~\ref{subsec:ResultsMeanField}.

%
\section{Triadic relations in empirical networks}
\label{sec:analysis}
In this section, we put the developed theoretical framework to the test  for both static (subsection~\ref{subsec:staticprobabil}) and dynamic (subsection~\ref{subsec:dynamicproba}) properties of the triads in the three political networks described in section~\ref{sec:data}. Also the predictions of the mean-field approximation to the proposed model are compared to the observations (subsection~\ref{subsec:ResultsMeanField}).
\subsection{Static probabilities from a quasi-equilibrium approximation}
\label{subsec:staticprobabil}

Under the assumption that the states of any two triads are  independent, the probability $p_{\sigma}$ of finding the triad $T_{ijk}$ in a certain state $\sigma$ is  determined by  
\begin{equation}
p_{\sigma} = \frac
{g_{G,T}(E_{\sigma}) e^{-\beta E_{\sigma}}}
{\sum_{\sigma'} g_{G,T}(E_{\sigma'}) e^{-\beta E_{\sigma'}}} =  
g_{G,T}(E_{\sigma})
e^{-\beta E_{\sigma} - \ln Z_0} \; ,
\label{eq:TriadicProbability}
\end{equation}
where we have introduced the normalization factor
\begin{equation}
Z_0 \equiv \sum_{\sigma} g_{G,T}(E_{\sigma}) e^{-\beta E_{\sigma}} 
\label{eq:Z0}
\end{equation}
and the degeneracies $g_{G}(E_{\sigma})$ and $g_{T}(E_{\sigma})$ as listed in table \ref{table:Levels}. These degeneracies measure the number of possible combinations that give rise to a certain triad. In the absence of any preference with regard to the triadic state all energies $E_{\sigma}$ are identical and the $p_{\sigma}$ are proportional to the degeneracies. By measuring the occupation probability $p_{\sigma}$ for each of the ten triadic states and by counting the corresponding degeneracies, one can extract the triadic energies $E_{\sigma}$ in any given time window. This allows one to determine the  triadic states that are overpopulated  (low energies $E_{\sigma}$) and are underpopulated (high energies $E_{\sigma}$) relative to a random network.
 
In reality, the states of two triads that share an edge, are not completely independent and taking those correlations into account is a notoriously challenging problem. To our knowledge, there is no formalism dealing with structural balance and its extensions that takes these correlations into account.
In the forthcoming we first introduce a quantity that allows us to get a handle on the correlations in the studied network. We then proceed with extracting triadic energies $E_{\sigma}$ under the independence assumption of equation~(\ref{eq:TriadicProbability}) and discuss the results for the three empirical networks described in section~\ref{sec:data}.

\begin{figure}[htb]
\centering
\includegraphics[width=0.60\textwidth]{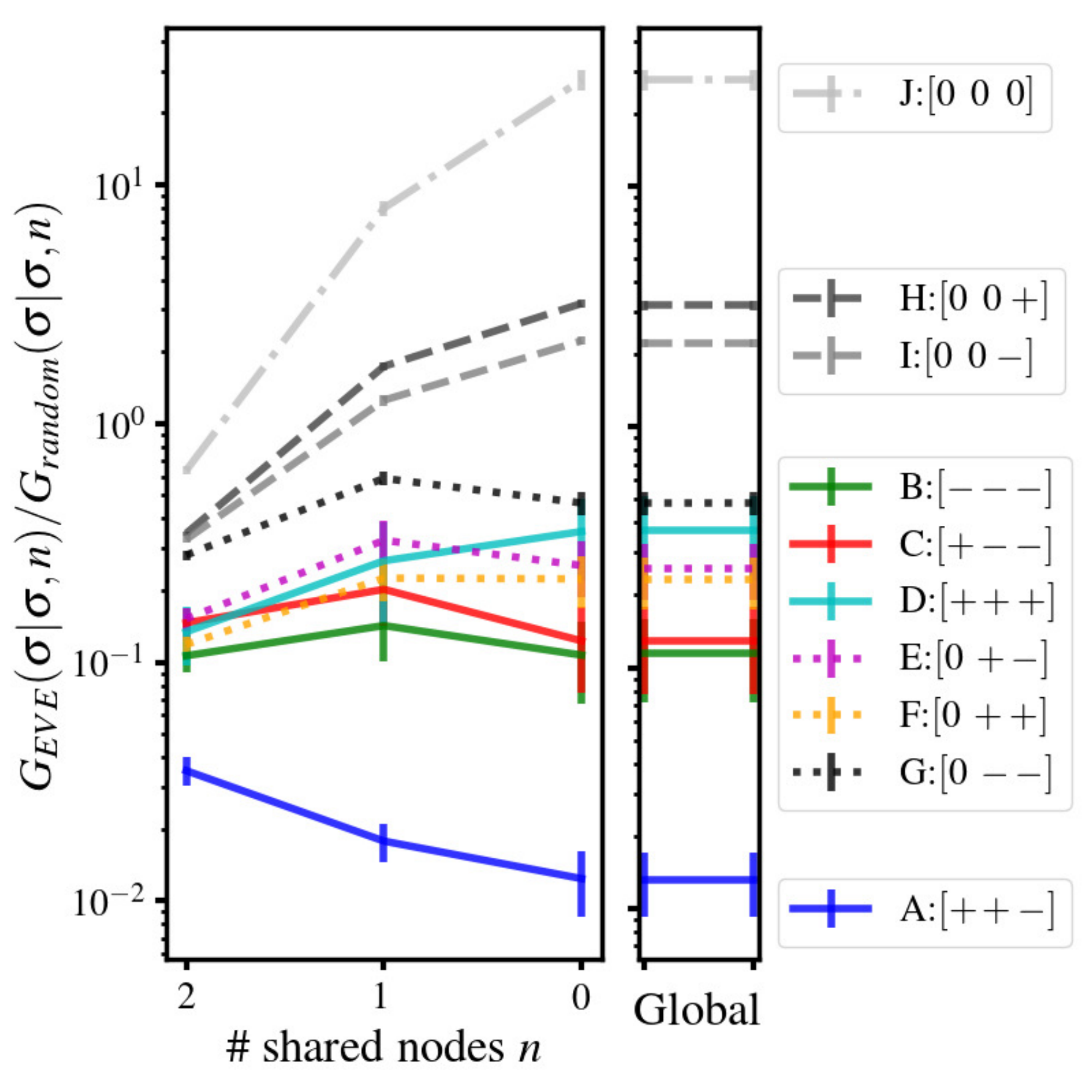}
\includegraphics[width=0.30\textwidth]{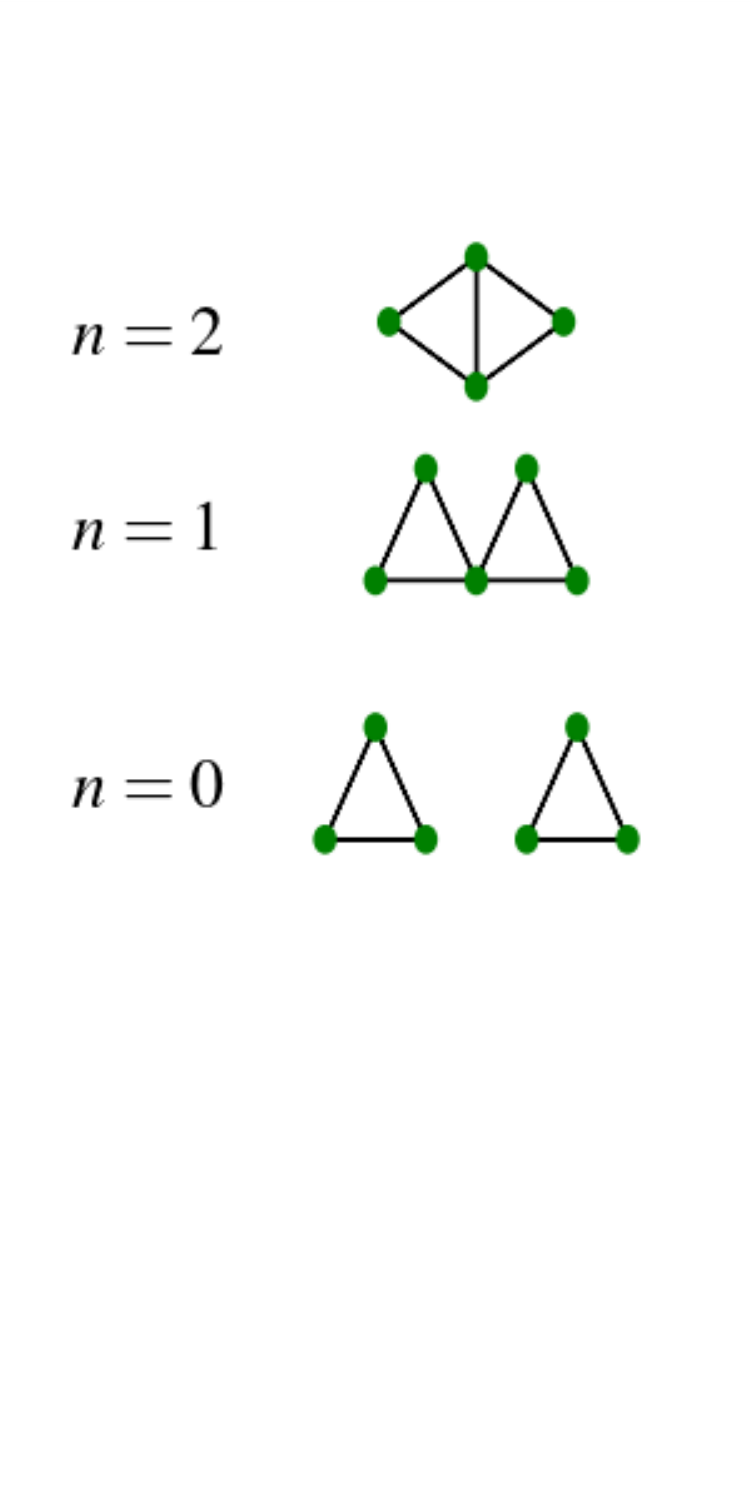}
  \caption{
  The conditional probabilities $G(\sigma |\sigma, n)$ in the EVE(SOV) dataset ($=G_{\textrm{EVE}}(\sigma |\sigma, n)$) relative to a random network ($=G_{\textrm{random}}(\sigma |\sigma, n)$). The $G(\sigma |\sigma, n)$ expresses the probability to find a triad in the state $\sigma$ sharing $n$ nodes with another triad in the state $\sigma$. We use a dot-dashed line for the triads with no activated edges, dashed line for triads with 1 activated edge, dotted line for triads with 2 activated edges, and  a solid line for the triads with 3 activated edges. 
 As a consistency check we also show the probability of finding a triad of type $\sigma$ without taking into account the triad's neighborhood.}

\label{fig:Correlation}
\end{figure}

As correlations can be anticipated to be strongest in the local neighborhood, we introduce a quantity that measures the correlations between the state $\sigma$ of a triad $T_{ijk}$ and the state of its neighbors. We define the probability of finding a triad of the type $\sigma$ given that it shares $n$ nodes with a triad of the type $\sigma ^{\prime}$ as $G(\sigma |\sigma', n)$. 
The $n$-dependence of $G(\sigma |\sigma, n)$ is instructive about the effect of triad clustering. Figure~\ref{fig:Correlation} displays $G(\sigma |\sigma, n=0,1,2)$ for the ten triadic states in the EVE(SOV) network  relative to the same quantity for a network where the same number of triads are randomly assigned.
For $n=0$ one retrieves the occupation probabilities of figure ~\ref{fig:Histogram_ocupation_probabilities} for the ten triadic states. We find that $G(\sigma |\sigma, n=1,2)$ significantly differs from $G(\sigma |\sigma, n=0)$. 
As the number of shared nodes increases, the values of $G(\sigma |\sigma, n)$ for all triads $\sigma$ tend to converge which alludes to a decreasing degree of correlations as one moves out of the local neighborhood. 

From figure ~\ref{fig:Correlation} we thus conclude that triads in the local neighborhood are not fully independent. 

Application of the equation~(\ref{eq:TriadicProbability}) to network data with few nodes is not fully justified. When zero nodes are shared, however, equation~(\ref{eq:TriadicProbability}) becomes more appropriate. With increasing system size, the proportion of local triads decreases and it can be anticipated that the validity of equation~(\ref{eq:TriadicProbability}) improves.

\begin{figure}[htb]
\centering
\includegraphics[trim={-0.5cm 2.6cm 0 0},clip,width=0.99\textwidth]{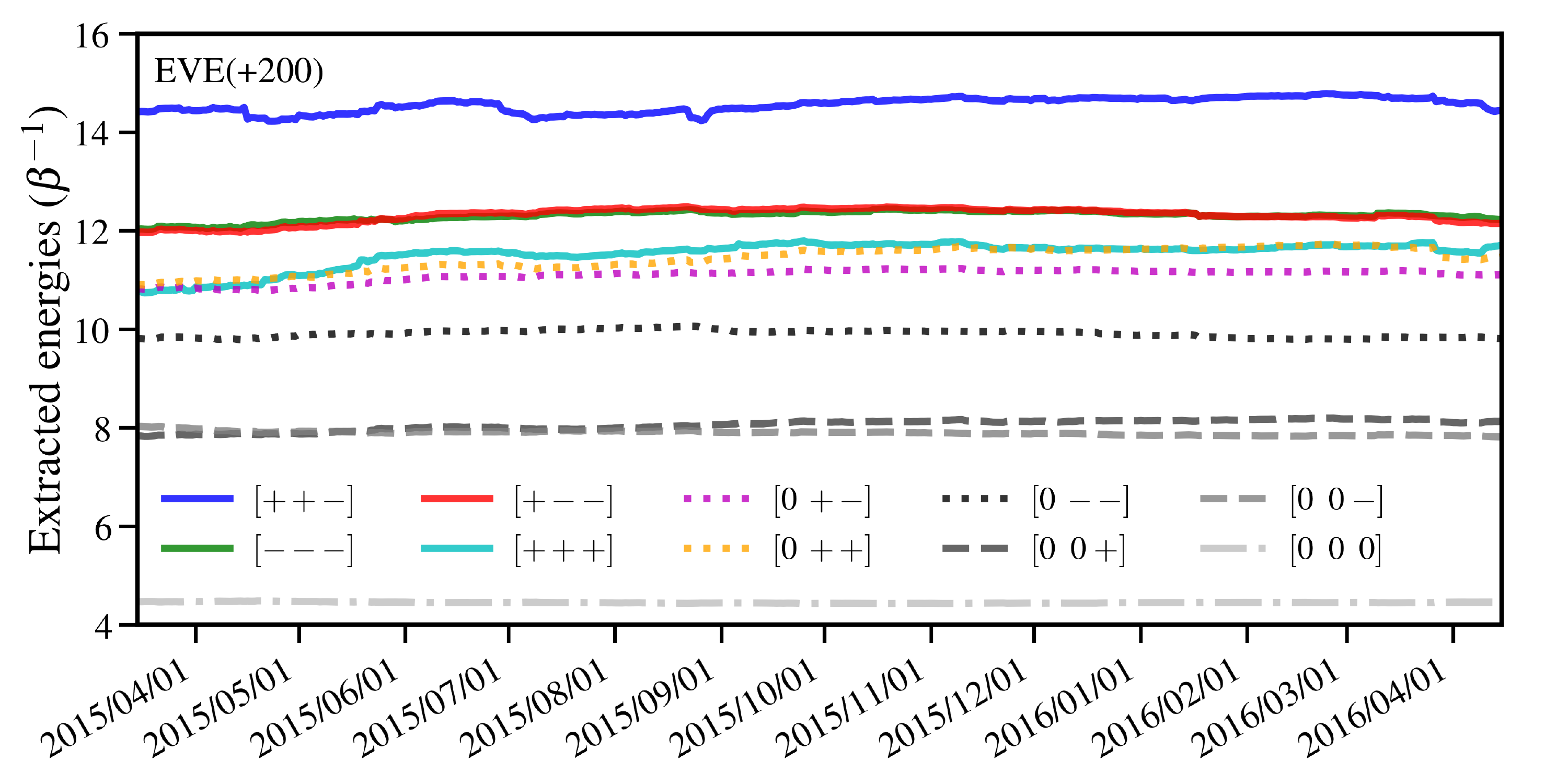}
\includegraphics[width=0.95\textwidth]{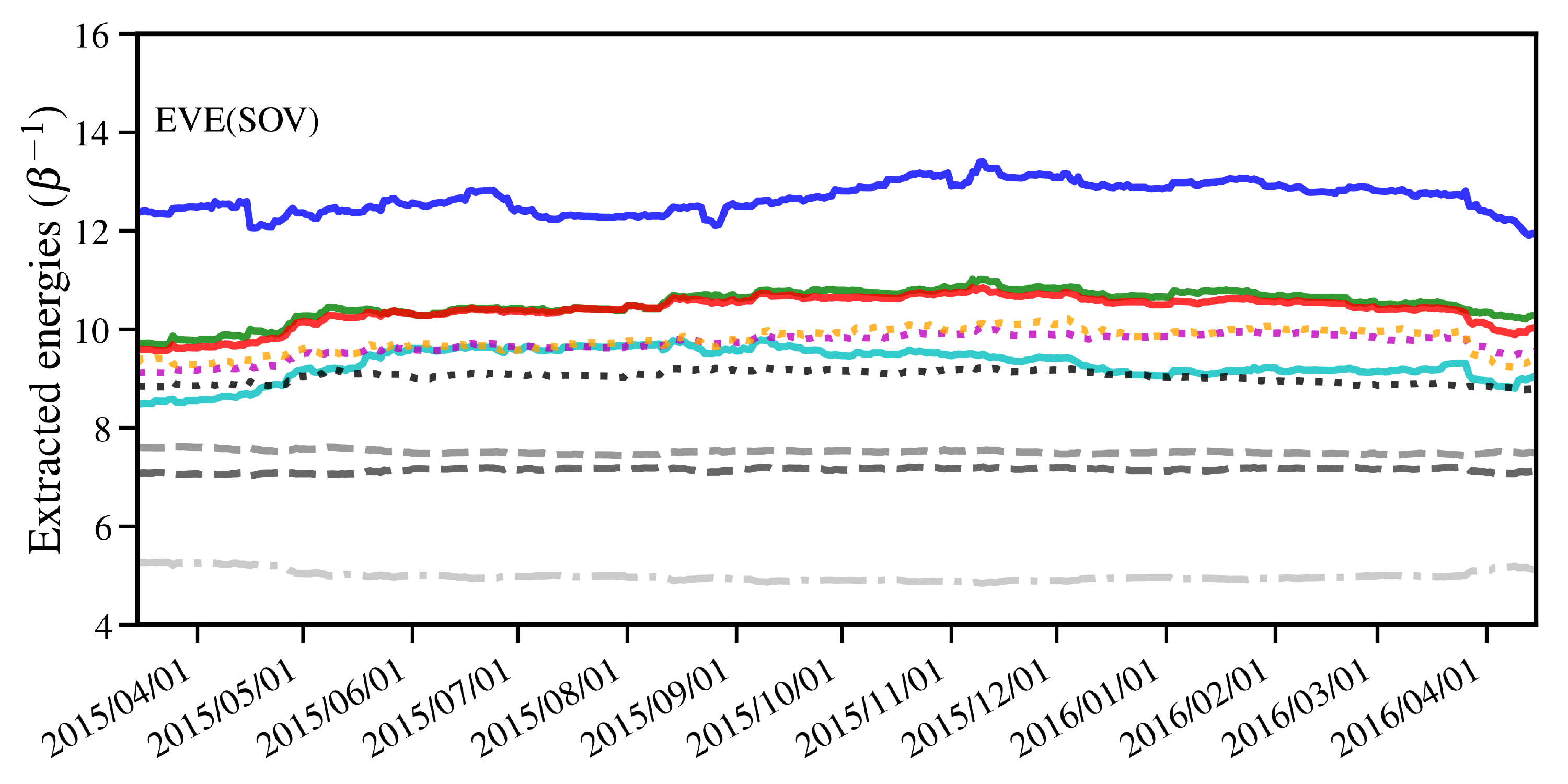}
\caption{ The logarithm of the daily occupation probabilities $-\ln p_{\sigma} /g_T (E_ \sigma)$ for the ten triadic states $\sigma$ for the {(top panel)} EVE(+200) and {(bottom panel)} EVE(SOV) of the alliances' network.
}

\label{fig:Normalized energy}
\end{figure}

\begin{figure}[htb]
\centering
\includegraphics[trim={0.5cm 0 0 0},clip, width=0.95\textwidth]{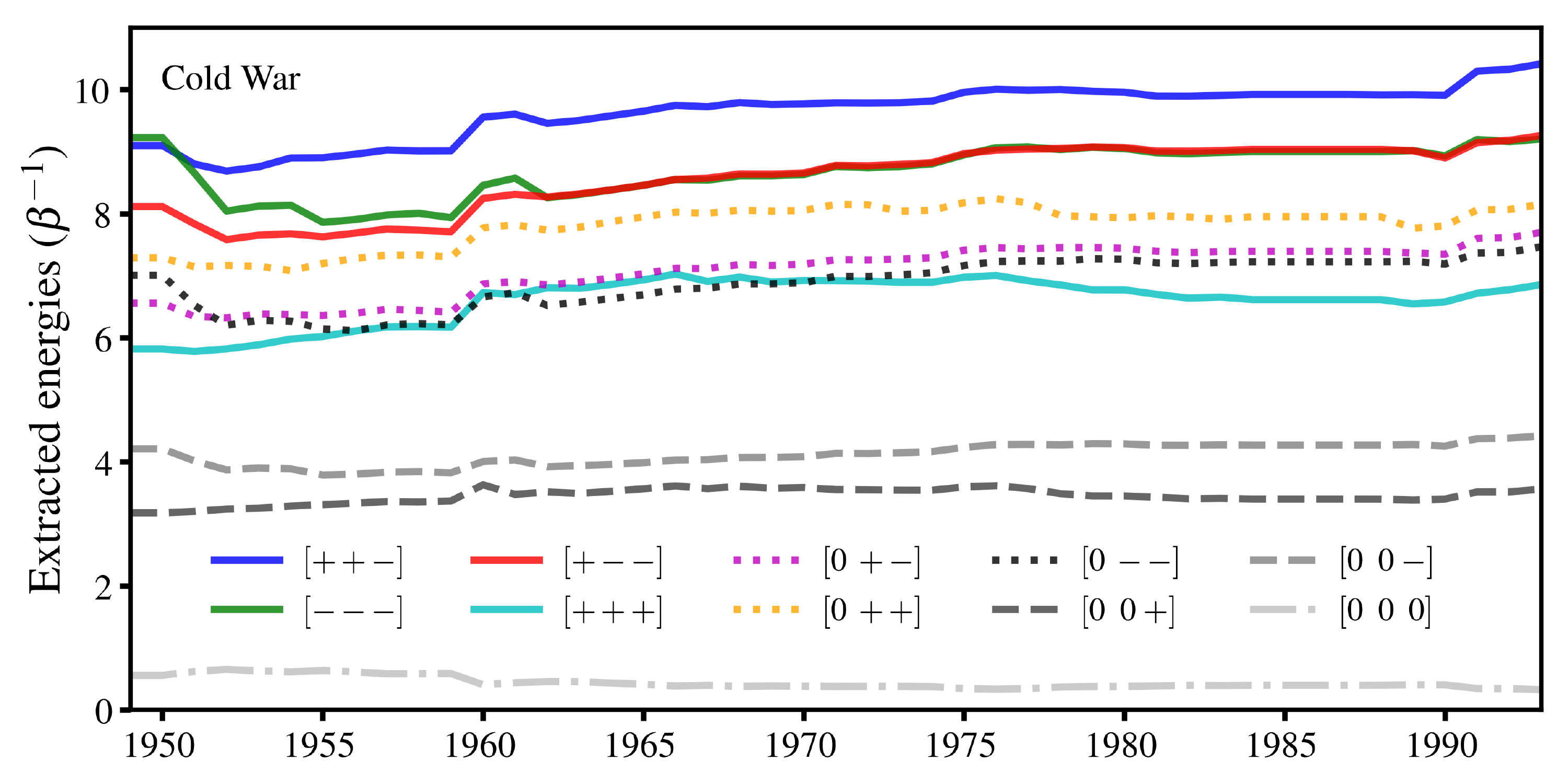}
\caption{The logarithm of the yearly occupation probabilities $-\ln p_{\sigma} /g_G (E_ \sigma)$ for the ten triadic states $\sigma$ of the inter-country relationships network during the Cold War era (1949-1993).
}
\label{fig:Normalized energy2}
\end{figure}

From the observed occupation probabilities $p_{\sigma}$, we can extract the energies $E_{\sigma}$ for each time window using~(\ref{eq:TriadicProbability}). The time series of the extracted $E_{\sigma}$ are displayed in figure~\ref{fig:Normalized energy} for the EVE data and in figure~\ref{fig:Normalized energy2} for the Cold War data. From figure~\ref{fig:Normalized energy} one can conclude the following: For both versions of the EVE network, one can discern a similar hierarchy in the relative position of the different types of triads.  First, unbalanced triads (dashed lines) have a higher energy than balanced triads (solid lines) which confirms the predictions of standard structural balance. Second, complete triads (non-gray colors) consistently have higher energies (less stable) than incomplete triads. Third, incomplete triads with only one inactive link are more energetic (less stable) than triads with two inactive links. This hierarchy inferred from the latter two observations suggests that the chemical potential is an essential term in the dynamics governing the formation of triads. As could already be inferred from figure~\ref{fig:Histogram_ocupation_probabilities}, there is clear separation in energy scale between the triads with zero and one active link. There is not a clear separation between the energies of the triads with two and three active edges. 

Furthermore, the close relative energy position for triads that are similarly affected by the chemical potential $\mu$ (e.g. $[0\;\;0\;+]$ and $[0\;\;0\;-]$), hints that $\omega$ is only a small perturbation. Finally, we note that the stability of the hierarchy of the energy levels over time suggests a roughly constant temperature.

As can be seen from figures~\ref{fig:Normalized energy} and \ref{fig:Normalized energy2}, the ordering of the extracted triadic energies is similar for the Cold War and EVE political networks. However, there are a few differences  for the complete triads. The energy gap for the strongly balanced $[+++]$ and the other three triads is higher for the real world than for EVE. Also, the energy separation between the other complete triads is smaller. Even so, these differences are rather small, so we expect similar values for the parameters. One also observes, as time progresses, a slight divergence between the higher and lower energy levels which hints at a cooling process in the political network of countries.

From the extracted energy levels, one can  infer the force strength parameters of the Hamiltonian $ \left\{ \alpha, \gamma, \omega \right\}$ and the value of the chemical potential $\mu$ using the relations contained in table~\ref{table:Levels}.
To this end, we use the time-averaged values $\overline{E_{\sigma}}$ with the standard deviation as a measure for the error.
Through a $\chi^2$ fit we find the parameters as listed in table~\ref{table:Adjusted Parameters1}. 
The results show a consistent hierarchy between the values of $ \left\{ \alpha, \gamma, \omega ,\mu \right\}$ with the chemical potential as dominant term. In line with the underlying ideas of structural balance  and with the results obtained in \cite{Belaza2017}, the strength parameter $\alpha$ of the three-edge force is larger than the strength parameter $\gamma$ of the two-edge force. The positive value of $\mu$ reveals that there is a substantial cost to establishing a relationship between two agents. The sizable positive value of the two-edge term $\gamma$ indicates that there are strong tendencies for homogenization in the three-node cycles.  For the EVE data, the extracted values for $\omega$ are small. This alludes to the fact that in EVE there are comparable incentives for ``$+$'' and ``$-$'' relationships between the alliances of players.  
It is remarkable, however, that the extracted parameters $ \left\{ \alpha, \gamma, \omega, \mu \right\}$  for all three studied networks are of the same order of magnitude. This gives rise to a rather universal hierarchy of the triadic energy states as it is shown in figure~\ref{fig:Diagram} and could also be inferred from the occupation probabilities of 
figure~\ref{fig:Histogram_ocupation_probabilities}.

\begin{table}
\centering
\begin{tabular}{ l c c c c c }
\hline \hline
& $\alpha$ $(\beta ^{-1})$ & $\gamma$ $(\beta ^{-1})$ & $\omega$ $(\beta ^{-1})$ & $\mu $ $(\beta ^{-1})$ & $\ln(Z_0)$  \\  \hline
EVE (SOV)  & $0.76\pm 0.20$ & $0.43\pm 0.08$ & $-0.10 \pm 0.06$ & $2.08 \pm 0.06$ & $5.22 \pm 0.08$ \\             
EVE (+200) & $0.73\pm 0.14$ & $0.54\pm 0.06$ & $+0.06 \pm 0.05$	& $3.06\pm 0.05$  & $4.48 \pm 0.06$ \\  
Cold War   & $0.68\pm 0.23$ & $0.42\pm 0.09$ & $-0.09 \pm 0.07$ & $3.00 \pm 0.07$ & $0.57 \pm 0.11$ \\  
\hline \hline
\end{tabular} 
\caption{The extracted values for the strength parameters in the Hamiltonian~(\ref{eq:Hamiltonian}) for the +200 and SOV alliances in EVE Online, and for the international relations during the Cold War era. In a random network $\alpha,\gamma,\omega, \mu = 0 $}
\label{table:Adjusted Parameters1}
\end{table}
\subsection{Dynamic transition probabilities}
\label{subsec:dynamicproba}

Starting from equation~(\ref{eq:TriadicProbability}) one can predict the transition probabilities between the 10 triadic states listed in table~\ref{table:Levels} and displayed in figure~\ref{fig:Diagram}.
In equilibrium, the relative occupation probabilities between the different triadic states are maintained. The normalized single-edge transition probabilities 

for a triad to change from state $\sigma$ to state $\sigma ^{\prime}$  
define the $10 \times 10$ non-symmetric probability matrix $\mathcal{M}$
\begin{equation}
\mathcal{M}_{\sigma  \sigma ^{\prime}}  = \frac {g_ { \sigma \rightarrow \sigma ^{\prime} } \exp -\beta \left( E_{\sigma ^{\prime} } - E_{\sigma} \right) }
{ \sum _{\sigma ^{\prime \prime}} g_ { \sigma \rightarrow \sigma ^{\prime \prime} } \exp -\beta \left( E_{\sigma ^{\prime \prime} } - E_{\sigma} \right) } \; .
\label{eq:singleedgetransition}
\end{equation}
Here, $g_{ \sigma \rightarrow \sigma ^{\prime} }$ denotes the total number of micro-states identified as a $\sigma ^{\prime}$ triadic state that are accessible from $\sigma$ through changing exactly one of the attributes of the micro-state. Examples of attribute changes can be a sign change in one of the  directed links between EVE alliances, or a change in an alliance treaty that corresponds with war declaration in the real-world data of inter-country relationships. The $g_{ \sigma \rightarrow \sigma ^{\prime} }$ can be interpreted as the dynamical equivalent of the degeneracies $g_G$ and $g_T$ listed in table~\ref{table:Levels}. 

For a considerable amount of transitions in the undirected network of inter-country relationships during the Cold War era, one has $\mathcal{M}_{\sigma \sigma ^{\prime}} =0$  as the corresponding $g_{\sigma \rightarrow \sigma ^{\prime}} =0$.  This  captures the fact that some triadic transitions are forbidden by means of changing the value of a single edge  (e.g.~the transition from $[+++]$ to $[---]$). For the network of alliances in EVE the situation is different. Specifically, in determining the $g_ { \sigma \rightarrow \sigma ^{\prime} }$ we start from the raw data for the edge attributes with occasional asymmetrically directed edges between two alliances in EVE.

\begin{figure}[htb]
\centering
\includegraphics[trim={0 0 2.0cm 0},clip,width=0.49\textwidth]{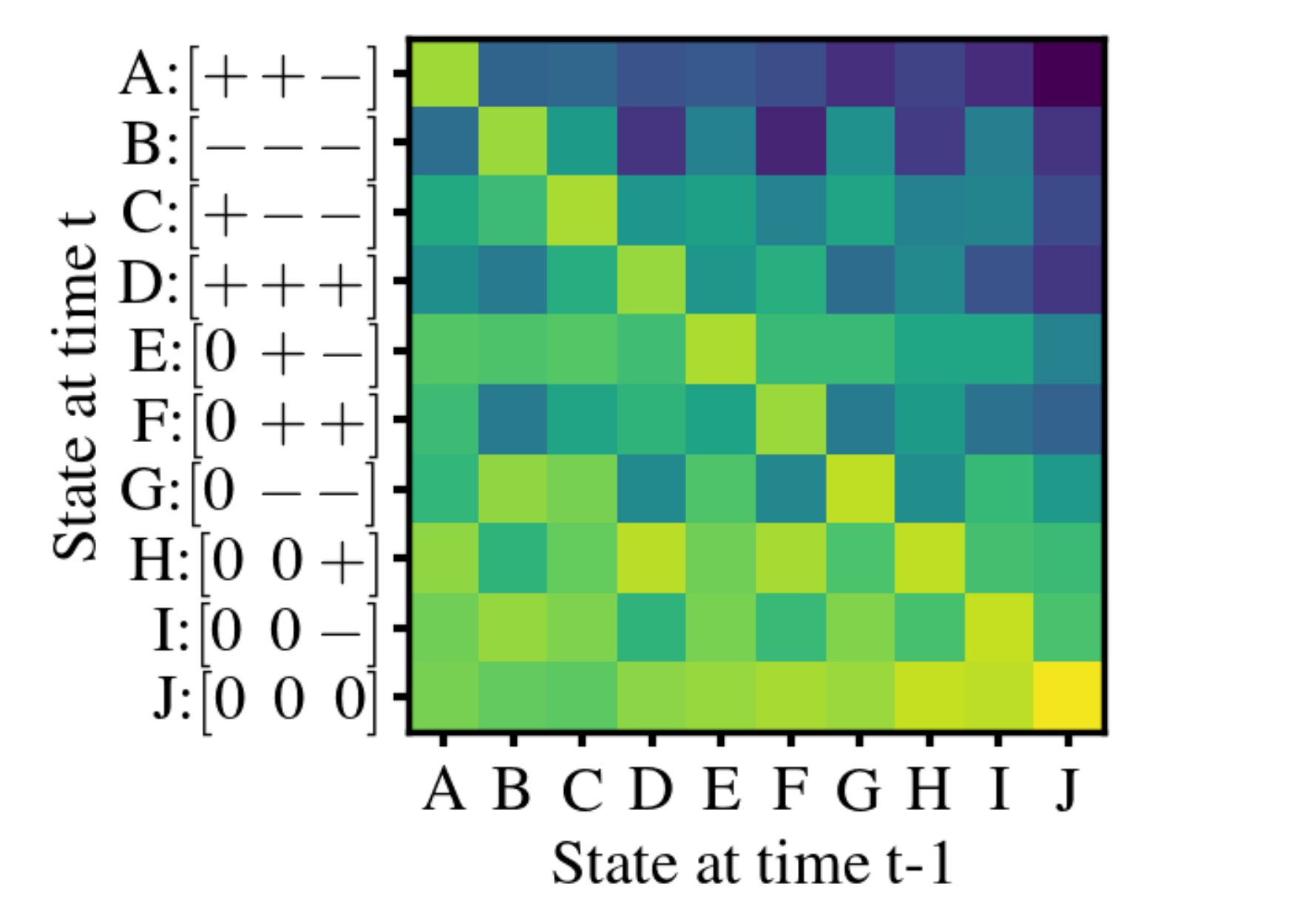}
\includegraphics[trim={2.0cm 0 0 0},clip,width=0.49\textwidth]{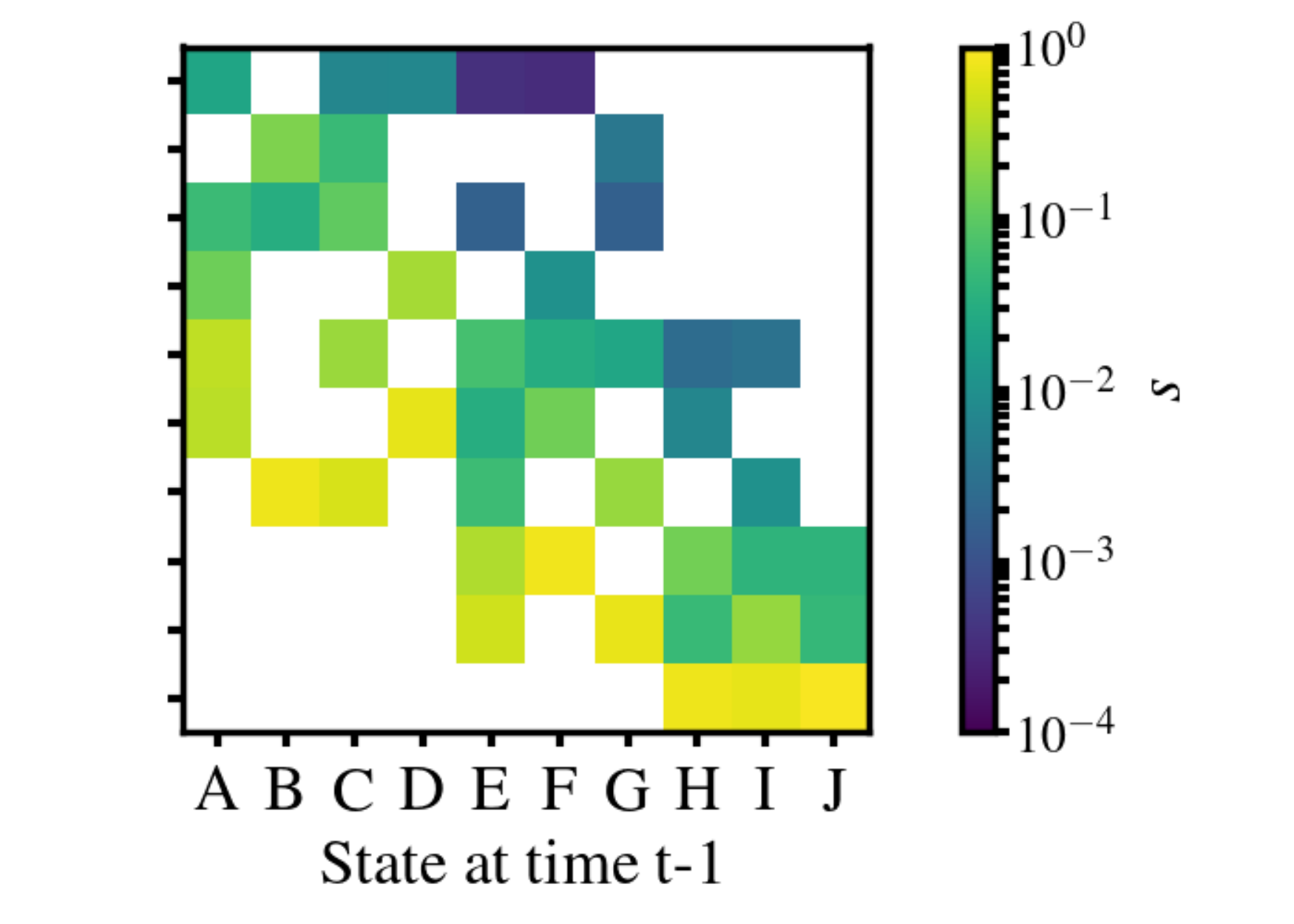}
\caption{The matrix $\mathcal{M}_{\sigma^{\prime} \sigma}$ 
for the intra-day single-edge transition probabilities between the ten triadic states in the EVE(+200) network. Left panel: the recorded transitions from April 9 to April 10, 2015.  Right panel: the predicted single-edge transitions of equation~(\ref{eq:singleedgetransition}). White sites correspond with $\mathcal{M}_{\sigma \sigma ^{\prime}} =0$.}
\label{fig:Transitions}
\end{figure}

In figure~\ref{fig:Transitions} we compare the predicted probability matrix $\mathcal{M}$ for the network of EVE(+200) alliances with the recorded transitions on April 9-10, 2015. This choice is inspired by the fact that on that particular day all possible single-edge transitions are recorded. We stress that some of those intra-day transitions are very infrequent. To compare the data with the model, we use (\ref{eq:singleedgetransition}) in combination with the extracted values for the energy values $E_{\sigma}$ (see figure~\ref{fig:Normalized energy}). 
The fact that the $g_ { \sigma \rightarrow \sigma ^{\prime} }$ are computed with the aid of directed networks,   explains why the diagonal $g_ { \sigma \rightarrow \sigma  } $ are non-vanishing, resulting in large diagonal $p_{\sigma \rightarrow \sigma}$. For example, the combinations $\left({s}_{1 \rightarrow 2}=+1, s_{2 \rightarrow 1}=0 \right)$ and $\left({s}_{1 \rightarrow 2}=+1, s_{2 \rightarrow 1}= +1 \right)$ both give rise to $s_{12}=+1$. The event of agent 2 changing its attitude towards agent 1 (denoted as $s_{2 \rightarrow 1}$) from ``$0$'' to ``$+$'' is recorded as a transition. Nevertheless, the initial and final edge value $s_{12}=+1$ in the undirected network did not change. 

We find that our model for the transition probabilities has predictive power in that the transitions above the diagonal (that correspond to transitions to states with a higher energy) are less common than those below the diagonal. The recorded and computed values of $\mathcal{M}_{\sigma  \sigma ^{\prime}}$ extend over several orders of magnitude. In the recorded daily transition data  we find ample transitions with a finite transition probability whereas the corresponding single-step transition probability $\mathcal{M}_{\sigma \sigma ^{\prime}} =0$. This discrepancy is a clear indication that multiple changes of a triad's state can occur during the measurement interval of a single day.  Hence, we face a situation where the time scale of the model and of the data recording is different. As a matter of fact, an extension of the single-edge triad transitions of equation~(\ref{eq:singleedgetransition}) to multiple-edge triad transitions is in order.


Let us define $W(n ; \Delta t)$ as the probability that an edge is subjected to $n$ updates in the time-measurement interval $\Delta t$. This allows us to write the multiple-edge transition probability $P _{\sigma \rightarrow \sigma^{\prime}} (\Delta t)$ to move from triadic state $\sigma $ to triadic state $\sigma^{\prime}$ in an arbitrary number of edge changes during the time interval $\Delta t$ as:   
\begin{eqnarray}
P _{\sigma\rightarrow \sigma^{\prime}} (  \Delta t ) & = & 
W(0;\Delta t) \delta _{\sigma, \sigma^{\prime}} \; + \;  
W(1;\Delta t) \mathcal{M} _{\sigma  \sigma^{\prime}} \; + \;  
W(2;\Delta t) \sum_i 
\mathcal{M} _{\sigma  i} \;
\mathcal{M}_{i  \sigma^{\prime}} \nonumber \\ 
& + & 
W(3;\Delta t) \sum_{i,j} 
\mathcal{M}_{\sigma  i} \;
\mathcal{M}_{ij} \;
\mathcal{M}_{j \sigma^{\prime}} + \cdots .
\label{eq:transitionManySteps}
\end{eqnarray}
For brevity of notation, we drop the explicit dependence of $W(n ; \Delta t)$ and $P_{ \sigma\rightarrow \sigma^{\prime}} (\Delta t )$ on $\Delta t$. 
In order to normalize the multiple-edge transition probabilities $P _ {\sigma\rightarrow \sigma^{\prime}} $ of equation~(\ref{eq:transitionManySteps}), we make use of the fact that the single-edge transition probabilities of equation~(\ref{eq:singleedgetransition}) are normalized: $\sum_{\sigma^{\prime}} \mathcal{M}_{k  \sigma^{\prime}} =1, \forall k$. This implies that
\begin{eqnarray}
\sum_{\sigma^{\prime}} P _{\sigma \rightarrow \sigma^{\prime}} & = & \sum_{\sigma^{\prime}} \biggl[ W(0) \delta _{\sigma, \sigma^{\prime}} 
+ W(1) \mathcal{M} _{\sigma  \sigma^{\prime}} + W(2) 
\sum_i \mathcal{M}_{\sigma  i} \; \mathcal{M}_{i  \sigma^{\prime}} 
\nonumber \\
& & +
W(3) \sum_{i,j} \mathcal{M}_{\sigma i} \; \mathcal{M}_{i j} \;  \mathcal{M}_{j  \sigma^{\prime}}
+...\biggr]
\nonumber \\
& = & W(0)  + W(1) + W(2) + W (3) + \ldots \; .
\end{eqnarray}

This means that the multiple-edge transition probabilities $P _{\sigma \rightarrow \sigma^{\prime}}$ can be normalized under the condition that
\begin{equation}
\sum_{n \geq 0} W(n) = 1 \; .
\end{equation}
Assuming that $W(n \geq 1) = \left[ W(1) \right]^n$ one obtains an expression that connects $W(0)$ to $W(1)$
\begin{equation}
1= W(0) + \sum_{n \geq 1} \left[ W(1) \right] ^n = W(0) + \frac{W(1)}{1-W(1)} \; .
\end{equation}
As the $W(n)$ are probabilities, the required convergence criteria are met. 
As a matter of fact, as $W(0)$ is a probability one has $0 \leq W(n) < \frac{1} {2}$.

In view of our above derivations, our final expression for the multiple-edge transition probability between the triadic states $\sigma$ to $\sigma ^{\prime} $ in the time interval $\Delta t$ reads
\begin{equation}
P _{\sigma \rightarrow \sigma^{\prime} }(\Delta t) = 
\frac { 1 - 2 W (1 ; \Delta t)} { 1 - W (1 ; \Delta t)} \delta _{\sigma, \sigma^{\prime}} 
\; + \; 
\left[ \sum_ {n \ge 1}  W^n (1 ; \Delta t) \mathcal{M}^n \right] _{\sigma \sigma^{\prime} } \;. 
\label{eq:totalTransProb}
\end{equation}

\begin{figure}[htb]
\begin{flushleft}
\centering
\includegraphics[trim={0 0 0 0},clip,width=1.0\textwidth]{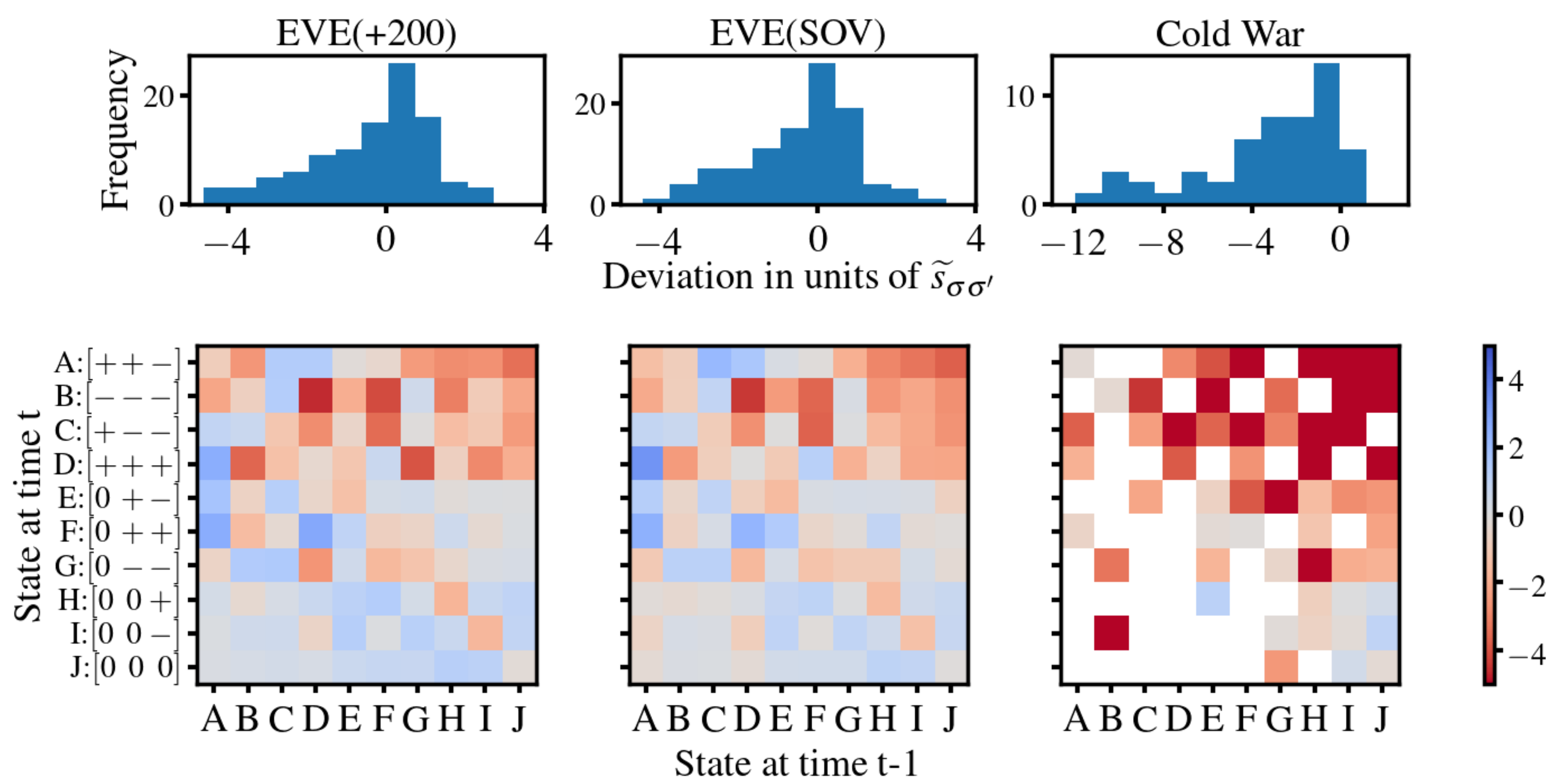}
\end{flushleft} 
\caption{The quality of agreement (quantified according to the equation~(\ref{eq:quality_of_agreement_transition})) between the predicted and the observed multiple-edge transition probabilities $P _{\sigma \rightarrow \sigma^{\prime} }(\Delta t)$ between the ten triadic states. The difference is positive (negative) when the model overestimates (underestimates) the mean of the observed transition probabilities. Left panels: EVE(+200); middle panels: EVE(SOV); right panels: Cold War era.  In the upper row we show the histograms of the $\chi _{\sigma \sigma ^{\prime}}$ values. In the lower row we show the matrix with $\chi _{\sigma \sigma ^{\prime}}$ values whereby light colors correspond to transition probabilities that are well predicted by the model.}
\label{fig:Transition_Diferences}
\end{figure}

To determine the functional form of the probability $W(1 ;\Delta t)$ we note there should be a dependence on the temperature and the constraint $0 < W(1;\Delta t) < 0.5$.  We propose the expression $W(1 ;\Delta t) = 0.5\exp(-E_p/T)$ inspired by equation~(\ref{eq:TriadicProbability}) with a positive-valued  $E_p$.
In line with our intuition, this expression gives rise to a higher number of transitions for higher temperatures.

To test the accuracy of the proposed model, we compare the theoretical multiple-edge transition probabilities of equation~(\ref{eq:totalTransProb}) with the observed ones. 
Given a set of parameters $\left\{ \frac {\alpha} {\mu}, \frac {\gamma} {\mu}, \frac{\omega} {\mu} \right\}$, an energy-activation-cost ($E_p$) and a temperature ($\beta ^{-1}$), one can compute the transition probabilities of equation~(\ref{eq:totalTransProb}) for each time window (intra-day for the EVE data and intra-year for the Cold War data). We stress that diplomatic relationships between alliances in EVE are highly volatile, in the sense that multiple alliances can change their attitudes within a time frame of a few days. In contrast, the network of international relationships during the Cold War era is rather static and an analysis of the transition probabilities is really challenging. 
From the data described in section~\ref{sec:data} we have collected  the recorded transition probabilities for each of the 100 $\sigma \rightarrow \sigma ^ {\prime}$ combinations. From those samples we determined  the expectation value $\widetilde{\mu}_{\sigma \sigma ^ {\prime}}$ and the standard deviation $\widetilde{s}_{\sigma \sigma ^ {\prime}}$ of the logarithm of the transition probabilities. For every transition $\sigma \rightarrow \sigma ^ {\prime}$, the quality of agreement between the model of equation~(\ref{eq:totalTransProb}) and the data is quantified by means of the difference between the predicted and measured transition probability relative to the variance of the distribution of the recorded $\sigma \rightarrow \sigma ^{\prime}$ transitions 
\begin{equation}
\chi _{\sigma \sigma ^{\prime}} =  \frac { \log \left( P _{\sigma \rightarrow \sigma^{\prime} } (\Delta t) \right) - \widetilde{\mu}_{\sigma \sigma ^ {\prime}} } {\widetilde{s}_{\sigma \sigma ^ {\prime}}} \; .
\label{eq:quality_of_agreement_transition}
\end{equation}
Figure~\ref{fig:Transition_Diferences} shows the quality of agreement for the transition probabilities between the model and the data for the three networks considered in this work. For both versions of the EVE alliance network, the majority of the values for the difference between predicted and observed transition probabilities is smaller than $ 2\widetilde{s}_{\sigma \sigma ^ {\prime}}$. We also learn that the model underestimates the perseverance of the ground state, predicting a lower probability to stay in $[0\;0\;0]$ than observed in the data. The model underestimates some transitions from balanced to unbalanced triangles (e.g.~$[+++]$ to $[---]$). The edge attributes of the network of inter-country relationships during the Cold War era are far more static than for the EVE networks. Accordingly, the number of recorded transitions is relatively small and about 40\% of the possible triadic transitions are not even recorded during the half-century covered by our data. As for the EVE data, the model severely underestimates the transition from a triadic state with at most one active edge, to one with three active edges (upper right corner of the transition matrix). Remarkably, the reversed transitions (lower left corner of the transition matrix) are reasonably well predicted. This seems to indicate that there is an incentive to generate triads with active edges that is not captured by our Hamiltonian approach. This could be related to another common idea in political science: ``You're either with us, or against us''. In other words, once one has revealed a few standings it often implies one is forced to announce others. In the model proposed here, this principle could be translated in an extra cost to maintain unclosed triangles.

\begin{figure}
\begin{flushleft}
\centering
\includegraphics[width=0.49\textwidth]{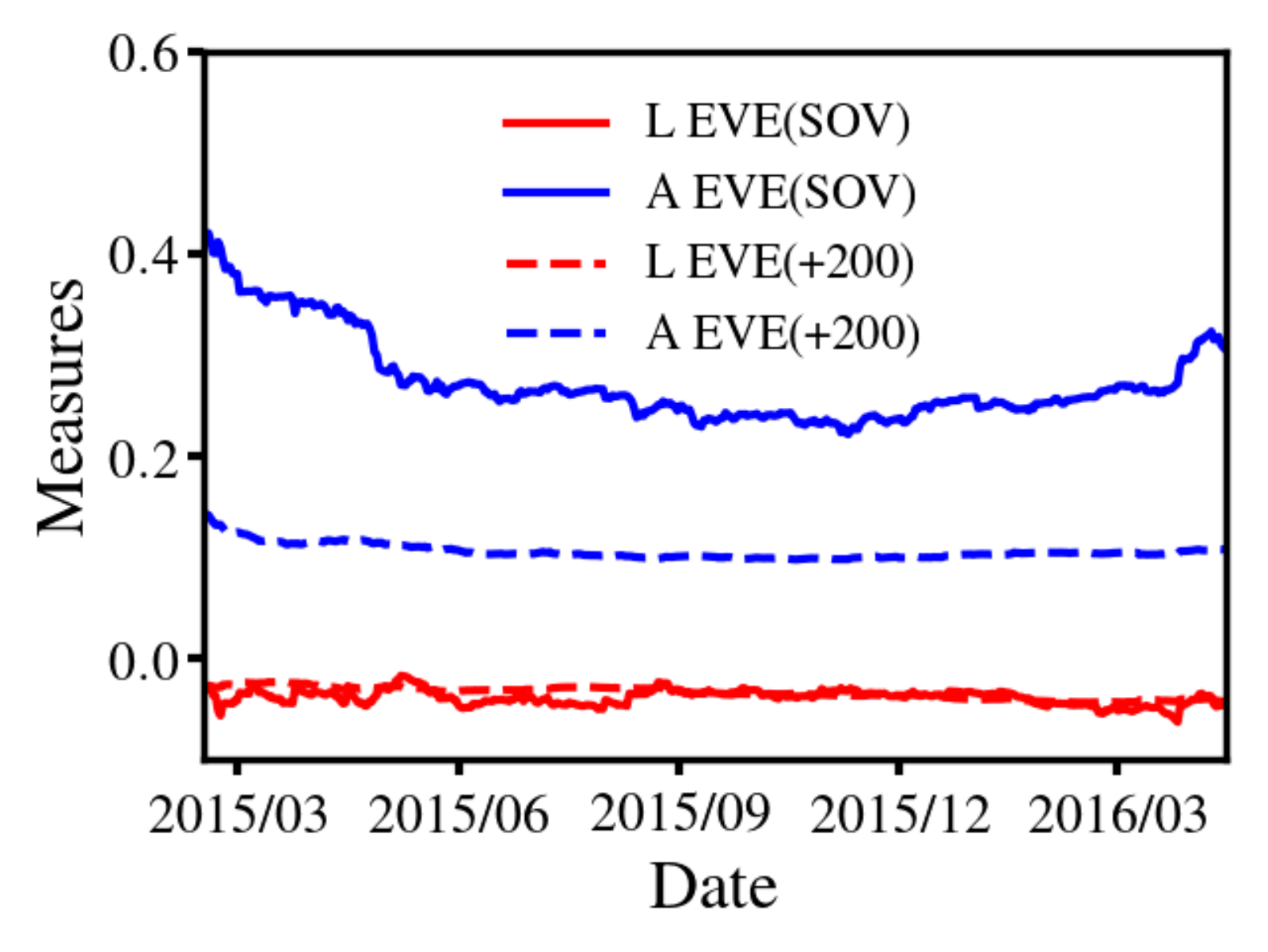}
\includegraphics[width=0.49\textwidth]{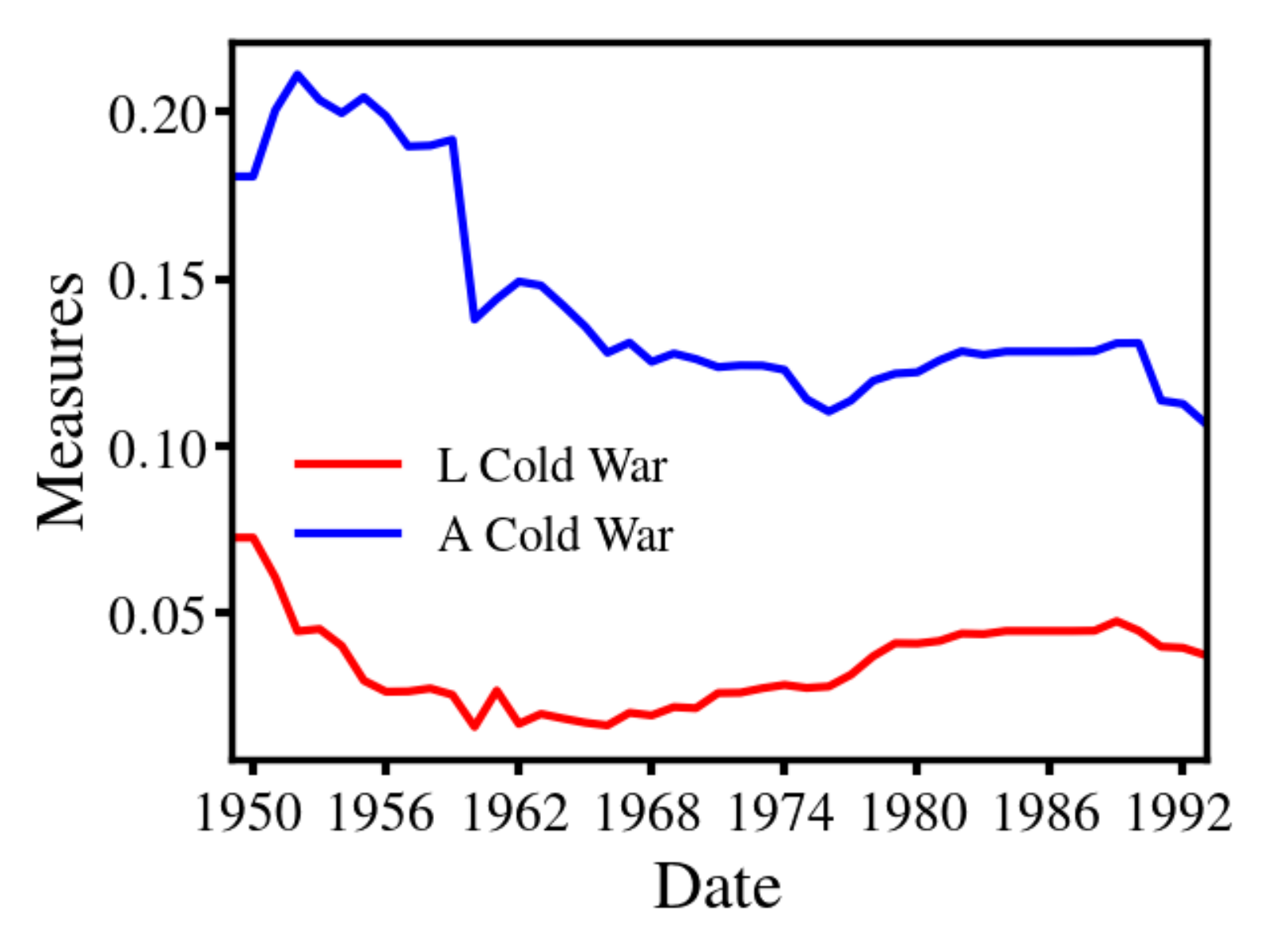}
\end{flushleft} 
\caption{Time series of the global measures $(A,L)$ for the networks of inter-alliance relationships in EVE (left) and of the inter-country relationships during the Cold War (right).}
\label{fig:TimeseriesGlobalMeasures}
\end{figure}

\subsection{Global measures}
\label{subsec:ResultsMeanField}

We now provide an analysis of the two global measures $L$ (average magnetization) and $A$ (average activation) and compare them with the mean-field predictions of section~\ref{subsec:MeanField}. 
Figure ~\ref{fig:TimeseriesGlobalMeasures} displays the time series of the extracted $L$ and $A$ for the three political networks considered in this work.  In the virtual world of EVE, $L$ is small and negative for both types of alliances. This indicates a small preference for a negative interaction between alliances in the virtual EVE world. The activation $A$ is markedly different for the two versions of EVE's alliance networks. For the network of large alliances EVE(+200), about 10\% of the links are active. For the alliances with sovereignty EVE(SOV), this is about 30\%. This substantial difference can be attributed to a size effect. We observe that the number of activated edges does not increase quadratically with the number of nodes.  As a result the fraction of active links decreases with the number of nodes in the network. Even with an infinite number of nodes, each node interacts with a finite number of other nodes.
For the Cold War network, the activation is subject to substantial time variations. Starting at around 20\% in the fifties, $A$ drops to 12-15\% in the sixties. In those days many countries  gained independence and it took them some time to create active ties. For a similar reason the collapse of the USSR (1990-1991) leads to a sudden drop in activation. In contrast to the EVE network, we observe a positive magnetization $L$.

The Eqs.~(\ref{eq:MFMagnetitation}) and (\ref{eq:MFExistence}) express $A = \left<s_{ij}^2\right>$ and $L = \left<s_{ij}\right>$ as a function of the temperature and two functions $c_2 = \mu$ and $c_1(N,L,\alpha,\gamma,\omega)$. 
From Eqs.~(\ref{eq:MFMagnetitation}) and (\ref{eq:MFExistence}) we can extract $ \beta c_1$ and $\beta c_2$ from given samples of the combination  $(L,A)$

\begin{eqnarray}
\beta c_1 & = & \textrm{arctanh} \left(\frac{-L}{A}\right) = \beta \left( \omega -\frac{2\gamma}{3} (N-2)L - \frac{\alpha}{3} (N-2)L^2  \right)\; ,
\label{eq:Betha_C1_from_l_A} \\
\beta c_2 & = & 
\ln \left(\left[\frac{1}{A} -1\right]2\cosh\left[\textrm{arctanh} \left(\frac{-L}{A}\right)\right] \right) = \beta \mu \; .
\label{eq:Betha_C2_from_l_A}
\end{eqnarray}

Upon close inspection of figure ~\ref{fig:TimeseriesGlobalMeasures} it is clear that $L$ is of the order of 0.1. As $N$ of the order of a few hundred, one has $(N-2)\cdot L \approx 1$. This implies that the ratio of Eqs.~(\ref{eq:Betha_C1_from_l_A}) and (\ref{eq:Betha_C2_from_l_A}) can be approximately written as  
\begin{equation}
\mathcal{G}_{MF} (L,A) \equiv \frac {\textrm{arctanh} \left(\frac{-L}{A}\right)} 
{\ln \left(\left[\frac{1}{A} -1\right]2\cosh\left[\textrm{arctanh} \left(\frac{-L}{A}\right)\right] \right)} =
\frac {\textrm{arctanh} \left(\frac{-L}{A}\right)} {\beta \mu} 
\approx \frac {\omega} {\mu} -\frac{2\gamma}{3 \mu} (N-2)L \; .
\label{eq:nontrivialrelation}
\end{equation}

\begin{figure}[htb]
\begin{flushleft}
\centering
\includegraphics[width=0.9\textwidth]{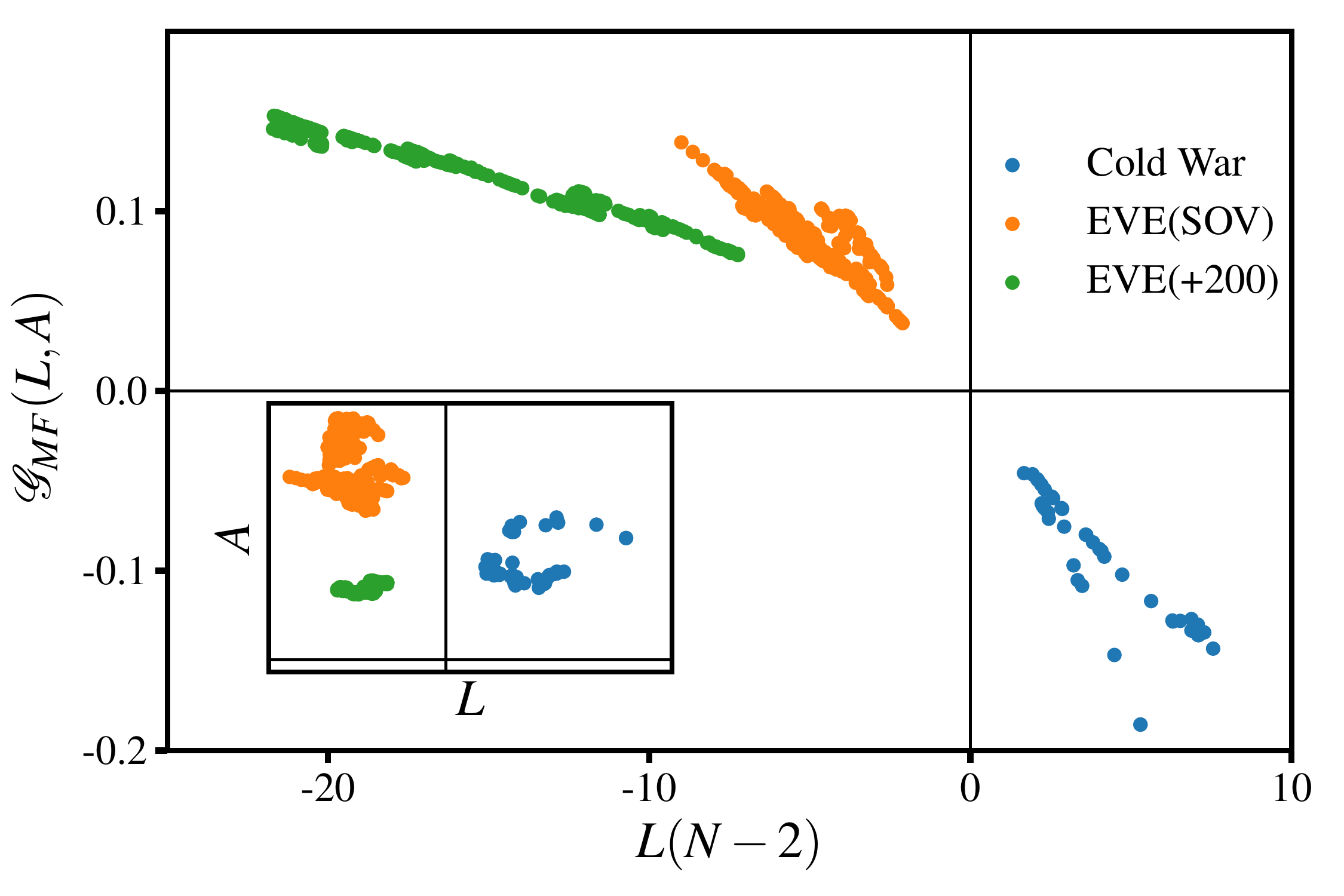}
\end{flushleft} 
\caption{Results of the empirical analysis of the predicted mean-field result~(\ref{eq:nontrivialrelation}) for the $(L,A)$ data of the EVE(SOV), EVE(+200) and Cold War networks. 
The inset shows the $(L,A)$ scatter plots. } 
\label{fig:MF-fit}
\end{figure}

As a matter of fact, this means that under conditions of constant parameters $ \{ \alpha, \gamma, \omega, \mu \}$, the mean-field model developed in section \ref{subsec:MeanField} predicts a linear relationship between the well-defined scaling function $\mathcal{G}_{MF} (L,A) $ and $L(N-2)$ for small values of $L$. 
The scaling function can be rewritten in a more convenient form 
\begin{equation}
\mathcal{G}_{MF} (L,A) \equiv \frac {\ln (1-\frac{L}{A}) -\ln (1+\frac{L}{A}) } 
{2\ln (2)+2\ln(\frac{1}{A}-1)- \ln (1-\frac{L^2}{A^2})} \; .
\label{eq:scaling_function}
\end{equation}
We evaluate this predicted non-trivial data collapse in figure ~\ref{fig:MF-fit} by plotting the l.h.s.~of equation~(\ref{eq:nontrivialrelation}) versus $L(N-2)$ for the data samples of $(L,A)$ collected from the time series of the three studied political networks.  The results  are encouraging. The values of $\mathcal{G}_{MF} (L,A)$ for the EVE(+200) network nicely cluster along a line. For the EVE(SOV) network two distinct lines are discerned. One of the lines of the EVE(SOV) data  concerns the first 100 days in the time series, whereas the second one refers to the rest of the data. Also the Cold War data cluster in two distinct lines separating the first decade of the data from the remainder.  Apparently, at the end of the 1950s-start of the 1960s, a series of events has induced changes in the inner workings of the network of inter-country relationships. This trend change coincides with the period between the Hungarian Revolution of 1956 and the construction of the Berlin wall (1961). The Hungarian Uprising of 1956 started the outflow of people out the East block that culminated with the surprise erection of the wall in 1961. Before this period, there was still some communist movement in Europe, and there were open lines and co-operation between the blocks. It could have still gone in the cooperative direction, but the violent response to the Budapest revolution and the construction of the Berlin wall destroyed this idea, creating a schism between the communist world and the western powers. After this, the world enters into a situation of clear blocks and every country was almost forced to affiliate itself with one or the other power. This new rule of ``if you are not a friend, you are an enemy''  increases the pressure of the triadic closure. This change in the dynamics gives rise to two different slopes in the scaling analysis of the data. The mean-field scaling function of equation~(\ref{eq:nontrivialrelation}) allows one  to detect changes in the way the international network rules itself. Indeed, it goes without saying that parameters ruling complex dynamic systems may fluctuate over time \cite{Mark2018}.
We mention that for both the EVE(SOV) and the Cold War data there are indications for the change points in the time series displayed in figure~\ref{fig:TimeseriesGlobalMeasures} but that the analysis of figure~\ref{fig:MF-fit} provides a clearer way to separate the different dynamical regimes. The scatter plot of $A$ versus $L$ in figure~\ref{fig:MF-fit} illustrates that the two global measures of the networks are not linearly correlated. In those scatter plots, however, the data for the three political networks tend to cluster in regions. For the Cold War and EVE(SOV) data two distinct clusters can be discerned.   

%

\section{Conclusion}
\label{sec:conclusion}

Standard structural balance builds on active (i.e. ``$+$'' and ``$-$'') edges and often succeeds in capturing the major features of the formation and evolution of alliances in political networks. Standard social balance, however, corresponds with a very small fraction (order of a single percent) of the triads in political networks which raises the question whether all information is being used. We have added the inactive edges and this has some clear advantages. Indeed, it introduces an additional layer of dynamics as active edges can change to inactive ones and vice versa. In addition, adding the inactive edge attribute allows one to analyze the political networks as complete networks. This facilitates a mean-field approach and the introduction of some representative global quantities of the network like the activation (fraction of active edges in the network) and the magnetization (average over edges states). 

 We have proposed a Hamiltonian that deals with positive, negative and neutral/nonexistent relationships as a representation of the generative dynamics in political networks. The Hamiltonian  contains single-edge, double-edge and triple-edge interactions inducing correlations in the edge attributes (``+'', ``$-$'' or ``0''). The Hamiltonian adds a three-edge interaction (that accounts for the major effect from structural balance) to the one from the Blume-Capel model. In a Boltzmann-Gibbs framework the model is suitable for theory-data comparisons and allows one to uncover and quantify the major dynamical mechanisms (including change points) at play in the political network.  We have put the Hamiltonian model as well as its mean-field approximation to the test with three extended sets of time-series data for the relationships between agents in political networks. Two sets of data are from the virtual world EVE Online and one set from the data about the inter-country relationships during the Cold War era.   Topics of investigation across the three networks considered have been the occupation probabilities of the ten triadic states as well as the transition probabilities among those states.  The occupation probabilities can be connected to the force terms in the Hamiltonian and some features are universal across the three network. We found that the cost of creating an active (hostile or friendly)  connection is higher than any other interaction component between the nodes. This locates the ground state of the system at a set of nodes with exclusively inactive connections in contrast to previous models which took universal utopia as the minimal energy system state. An empirical analysis of the three political networks revealed that about half of the possible triads reside in the $[0\;0\;0]$ state. There is a plethora of incentives (captured by the concept of temperature in our model) that result in agents interacting in hostile or friendly ways and causes the political system to reside in an excited state.  
  
With regard to the transition probabilities a fair model-data agreement is obtained for the EVE data where changes in the edge attributes are abundant.  In the Cold War data the edge attributes are rather static and the model does not describe the data very well. Across the three networks the model systematically underestimates a peculiar type of transition that is referred to as triadic closure. This corresponds with the transition from a triadic state with one active and two inactive edges, to a triadic state with exclusively active edges. This implies that upon activating one edge in the triad there are strong incentives to close the triad that are not captured by our current model.

The mean-field approximation for our model of political networks can be developed along the lines of the Blume-Capel model and results in a set of two self-consistent equations for the average magnetization $L$ and the average activation $A$ of the edges. Those two global variables allow one to sketch the phase boundaries between ordered and disordered realizations of the political networks. We find that the three studied networks fall in an intermediate partially ordered state. Thanks to the mean-field theory we could uncover a linear scaling between a non-trivial function of $(L,A)$ and $L$. An analysis of the empirical $(L,A)$ data provided evidence for the predicted  linear scaling. The mean-field prediction also turned out to be an efficient method for change-point detection and identify the different modes of operation as the political networks evolve over time.    

\section*{Acknowledgments}

This work was supported by Fund of Scientific Research - Flanders (FWO) (JR, KS, MvdH,BV) and Research Fund of Ghent University (AMB, KH). The funders had no role in study design, data collection, and analysis, decision to publish, or preparation of the manuscript.

\bibliographystyle{ieeetr}
\bibliography{bibliography}

\end{document}